\documentclass[iop]{emulateapj}   

\usepackage{graphicx}
\usepackage{epsfig}
\usepackage{amsmath}
\usepackage{subfigure} 
\usepackage{hyperref}
\usepackage{amssymb}
\usepackage{apjfonts}
\usepackage{latexsym}
\usepackage{color}
\usepackage[usenames,dvipsnames,svgnames,table]{xcolor}
\usepackage{hyperref}
\usepackage[all]{hypcap}    

\shorttitle{Pop III formation in cosmological volumes} 
\shortauthors{B.~D. Crosby et al.}

\begin{document}\title{Population III star formation in large cosmological volumes I.  Halo temporal and physical environment}

\author{Brian D. Crosby}\affil{Department of Physics and Astronomy, Michigan State University, East Lansing, MI 48824, USA}\email{crosbyb1@msu.edu}
\author{Brian W. O'Shea}\affil{Department of Physics and Astronomy, Michigan State University, East Lansing, MI 48824, USA}\affil{Institute for Cyber-Enabled Research, Michigan State University, East Lansing, MI 48824, USA}\affil{Lyman Briggs College, Michigan State University, East Lansing, MI 48825, USA}
\author{Britton D. Smith}\affil{Department of Physics and Astronomy, Michigan State University, East Lansing, MI 48824, USA}
\author{Matthew J. Turk}\affil{Department of Astronomy, Columbia University, New York, NY 10025, USA}
\author{Oliver Hahn}\affil{Institute for Astronomy, ETH Zurich, CH-8093 Z\"{u}rich, Switzerland}

\label{firstpage}

\begin{abstract}

We present a semi-analytic, computationally inexpensive model to identify halos capable of forming a Population III star in cosmological simulations across a wide range of times and environments.  This allows for a much more complete and representative set of Population III star forming halos to be constructed, which will lead to Population III star formation simulations that more accurately reflect the diversity of Population III stars, both in time and halo mass.  This model shows that Population III and chemically enriched stars coexist beyond the formation of the first generation of stars in a cosmological simulation until at least $z\sim10$, and likely beyond, though Population III stars form at rates that are 4-6 orders of magnitude lower than chemically enriched stars by $z=10$.  A catalog of more than 40,000 candidate Population III forming halos were identified, with formation times temporally ranging from $z=30$ to $z=10$, and ranging in mass from $2.3\times10^5~$M$_\odot$ to $1.2\times10^{10}~$M$_\odot$.  At early times, the environment that Population III stars form in is very similar to that of halos hosting chemically enriched star formation.  At later times Population III stars are found to form in low-density regions that are not yet chemically polluted due to a lack of previous star formation in the area.  Population III star forming halos become increasingly spatially isolated from one another at later times, and are generally closer to halos hosting chemically enriched star formation than to another halo hosting Population III star formation by $z\sim10$.

\end{abstract}


\section{Introduction}

Accurately modeling Population III star formation faces a myriad of challenges.  The inability to directly observe Population III stars requires recourse to theoretical investigations, and the intricacies of stellar formation and evolution make a detailed analytic approach intractable.  Population III stars form in dark matter halos that are not large relative to the star forming cloud, and properties of the pre-stellar cloud are related to the halo formation history \citep{2007ApJ...654...66O}.  Given the importance of halo formation history, primordial star formation must be studied in a cosmological context.  Numerical simulations have made great progress in our understanding of Population III stars, but finite computational resources require a tradeoff between cosmological volume and simulation resolution.  Properly modeling Population III star formation requires multiphysics simulations that have high spatial and temporal resolution, and by necessity simulate a small number of halos in small cosmological volumes typically of a few hundred kpc on a side (e.g., \citet{2002Sci...295...93A,2002ApJ...564...23B,2007ApJ...654...66O,2009Sci...325..601T}).  Studies of high redshift galaxy formation simulate larger volumes, but are limited to resolving only a single halo object (e.g., \citet{2002ApJ...575...49R,2012MNRAS.427..311W,2012ApJ...745...50W}) and are unable to resolve the intricacies of star formation.  These limitations, studying only small volumes and individual objects, provide very weak statistics on the nature of the Population III stellar population as a function of time and environment, an understanding of which is critical to determining the role of Population III stars in galaxy formation.  Smoothed particle hydrodynamics simulations have endeavored to constrain the Population III IMF, both with the use of sink particles (e.g., \citet{2012arXiv1211.1889S,2012MNRAS.422..290S,2011ApJ...727..110C,2011Sci...331.1040C}) and without \citep{2012MNRAS.424..399G}, the latter employing a moving-mesh method.  These works indicate that the Population III initial mass function  might be strongly dependent on the nature of protostellar accretion and disk fragmentation, processes which could potentially mitigate differences in the initial conditions of the pre-stellar cloud.

The statistical power of a simulation can be increased by simulating many Population III star forming halos simultaneously in larger cosmological volumes, but these studies have typically been limited to methods such as the extended Press-Schechter formalism \citep{2005ApJ...629..615W} in conjunction with a prescription for star formation \citep{2009ApJ...694..879T}.  While these models are capable of investigating some of the statistics of the population, they are unable to directly investigate the nature of individual Population III stars.  Advanced codes such as MUSIC \citep{2011MNRAS.415.2101H} and GRAFIC \citep{2001ApJS..137....1B} create initial conditions that can span a huge dynamic range, allowing for simulations of cosmological volumes that are still able to resolve, at a resolution appropriate for multiphysics simulations, individual halos that will host star primordial formation.  It is computationally prohibitive to simulate an entire cosmological volume of the desired size that also has the appropriate resolution to study star formation, presenting a new problem: which halos are likely to form primordial stars?  When a halo forms a star, later generations of star formation will be affected globally by the photodissociating and ionizing radiation produced by the star during its lifetime.  Later generations of star formation will be affected locally by metals produced by the star, which is likely to cause an abrupt change in the star-forming properties of the cloud (e.g., \citet{2001MNRAS.328..969B,2003Natur.425..812B,2009ApJ...691..441S}).  To study Population III stars with the detail required to investigate the stellar initial mass function (IMF) requires multiphysics simulations over a wide range of redshift and environment.  Halo merger history is crucial in the development of star forming clouds \citep{2002ApJ...564...23B,2007ApJ...654...66O}, as is the effect of photodissociating radiation \citep{2008ApJ...673...14O,2001ApJ...548..509M}.  These factors must be treated in a self-consistent manner in order to properly determine which halos will form Population III stars.  

This paper presents the first results from a model that seeks to explore the bulk properties of primordial star-forming halos over a large range of redshifts, and in much larger cosmological volumes than are typically used in simulations of Population III star formation.  In this model we account for halo merger history, metal enrichment and the transition to non-primordial star formation, and the effect of photodissociating radiation produced by all populations of stars.  Previous studies have typically ensured that the halos selected to host a Population III star are chemically pristine and unaffected by photodissociating radiation by choosing the first, most massive halo in the simulation \citep{2002Sci...295...93A,2007ApJ...654...66O}, though these objects are inherently rare and not necessarily representative of all Population III star forming halos.  The Population III IMF is poorly constrained theoretically, with estimates ranging from tens of solar masses \citep{2008ApJ...681..771M,2007MNRAS.382..229S} to hundreds of solar masses \citep{2002Sci...295...93A,2007ApJ...654...66O,2004ApJ...603..383T}.  Observational constraints suggest that the mean mass of primordial stars is on the order of tens of solar masses \citep{2006ApJ...641....1T}, which is in agreement with some cosmological simulations of Population III star formation (e.g., \citet{2009Sci...325..601T}).  The absence of any observed primordial stars presents circumstantial evidence that sub-solar mass Population III stars are very rare.  It is unclear to what extent the current theoretical predictions are a result of selection effects from cosmological simulations, and a more representative sample of primordial star-forming halos may provide us with a more realistic range of predictions.  This work presents a semi-analytical model of structure formation that can be applied to an \emph{N}-body simulation to determine which halos are likely to be chemically pristine and capable of hosting a Population III star, given the previous history of the halo and the global radiation background provided by stars forming elsewhere in the simulation volume.  Given the tools that were used to generate the initial conditions for these simulations, it will be straightforward to then resimulate individual halos of interest at high resolution and with more physics, including hydrodynamics, a full non-equilibrium primordial chemistry network, radiative cooling, and the effect of photodissociating background radiation.

In this paper, the first of a series, we describe in detail the model we use and its behavior, and validate it using multiple observational constraints.  We use this model to investigate the bulk behavior of primordial and metal-enriched stellar populations as a function of redshift, halo mass, and environment.  The organization of this paper is as follows: the cosmological simulations used are discussed in \S\ref{sec:simulations}, our semi-analytic model is described in \S\ref{sec:model}, and results are presented in \S\ref{sec:results} and discussed in comparison to observations and other computational work and future work in \S\ref{sec:discussion}.  We summarize our findings in \S\ref{sec:conclusions}.

\section{Simulations}
\label{sec:simulations}

\subsection{Enzo}
The simulations used as a basis for our model were carried out using the publicly available Enzo\footnote{http://enzo-project.org}  adaptive mesh refinement + \emph{N}-body code \citep{1997ASPC..123..363B,2000IMA...117..165B,1999ASSL..240...19N,2004astro.ph..3044O,2005ApJS..160....1O}.  Four simulations were run: two with a comoving box size of $3.5~h^{-1}$Mpc and two with a comoving box size of $7.0~h^{-1}$Mpc.  We used two different simulation volumes and two random realizations per chosen volume to give some idea of the impact of cosmic variance as well as mass and spatial resolution on our results.  We use the WMAP 7 best-fit cosmological model \citep{2011ApJS..192...18K}, with $\Omega_{\Lambda}=0.7274$, $\Omega_{M}=0.2726$, $\Omega_{B}=0.0456$, $\sigma_8=0.809$, $n_{s}=0.963$, and $h=0.704$ in units of $100~$km s$^{-1}$Mpc$^{-1}$, with the variables having their usual definitions.  All simulations are cubic and have 1024 grid cells per edge and $1024^3$ dark matter particles, giving cell dimensions of $6.8~h^{-1}$ comoving kpc on a side, a dark matter particle mass of $2.86\times10^4~$M$_\odot$, and a mean baryonic mass per cell of $5.74\times10^3~$M$_\odot$ for the $7.0~h^{-1}$Mpc boxes.  The $3.5~h^{-1}$Mpc boxes have cell dimensions of $3.4~h^{-1}$ comoving kpc on a side, a dark matter particle mass of $3.57\times10^3~$M$_\odot$, and a mean baryonic mass per cell of $718~$M$_\odot$.  

The simulations were initialized at $z=99$ using the MUSIC cosmological initial conditions generator \citep{2011MNRAS.415.2101H} with a second-order Lagrangian perturbation theory method and separate transfer functions for dark matter and baryons.  A second-order Lagrangian perturbation method is necessary to obtain converged halo mass functions at such early times and high redshifts as the start of Population III star formation.  Each of the sets of initial conditions were generated using a different random seed.  The simulations were run with Enzo's unigrid (non-adaptive mesh refinement) mode and with adiabatic hydrodynamics from $z=99$ to $z=10$.  Data is output at integer redshifts until $z=14$, at which point the elapsed time between integer redshifts would exceed the timescale for star formation.  After $z=14$, data is output every $11~$Myr.  The simulation is stopped at $z=10$ to prevent modes on the order of the size of the simulation volume from entering the non-linear regime.

\subsection{Halo Finding and Merger Tree Creation}
Halos for all data outputs in the simulations were identified using the Friends-of-Friends (FOF) \citep{1985ApJS...57..241E} halo finder implemented in the yt\footnote{http://yt-project.org} analysis toolkit \citep{2011ApJS..192....9T} with a linking length of $0.2$ times the mean interparticle spacing.  Once the halos were found at each redshift, it was necessary to determine their merger history.  Halos are considered to have merged when particles from two or more separately-identified halos from the previous data output are present in a single halo at the current time.  yt was used to create a merger tree that traced the history of each halo based on the halos that had previously merged together to form it.  All of the halos that merge together to form a halo between two consecutive data outputs are considered to be "parent" halos, with the resulting halo at the current data output denoted as the "child" halo.  A child halo can have an arbitrary number of parents, and the possibility of halo fragmentation at later times is also considered, allowing a parent halo to break into multiple child halos.  In addition to following the lineage of halos, the percentage of the mass of a child halo that comes from each parent is also tracked, allowing for the complete accounting of dark matter mass in halos.

\begin{figure}[htbp] 
   \centering
   \includegraphics[width=.45\textwidth, clip=true]{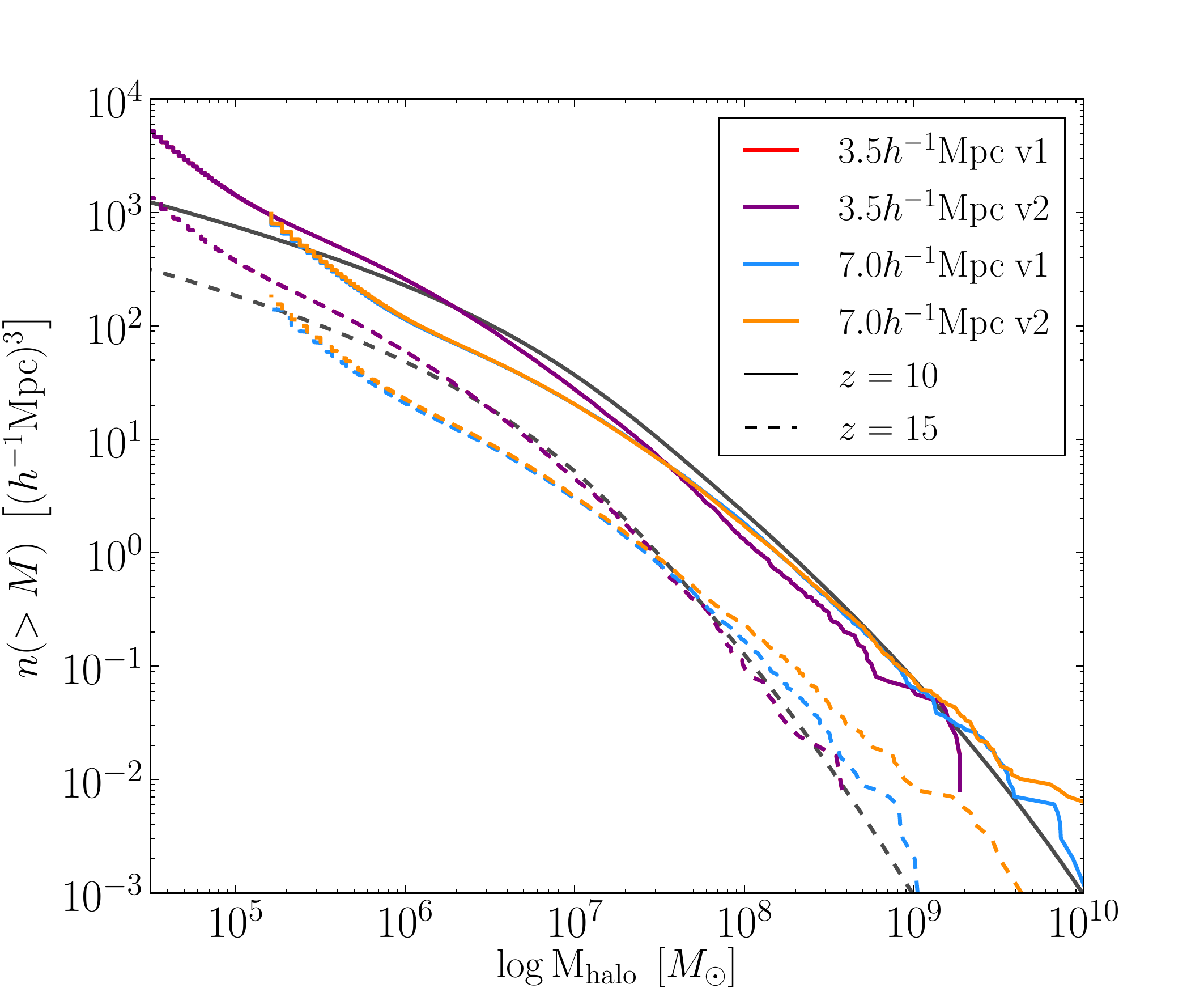} 
   \caption{The cumulative halo dark matter mass function for all four simulation volumes at $z=15$ (dashed line) and $z=10$ (solid line).  Plotted in black is the analytic prediction from the Warren mass function \citep{2006ApJ...646..881W}.  There is good agreement between the halo mass functions in the simulations over the redshifts of interest in this project.}
   \label{fig:mass_function}
\end{figure}

Figure \ref{fig:mass_function} shows the cumulative halo dark matter mass functions of all 4 simulations at $z=15$ and $z=10$, plotted along with the analytic Warren mass function \citep{2006ApJ...646..881W}.  The halo mass function in the simulations began to deviate from the analytic mass function below a dark matter mass of $2\times10^6~$M$_\odot$. The implementation of FOF that was used only takes into account the spatial distribution of particles, and will create a halo out of any group of 8 more particles that are sufficiently spatially clustered, without regard for whether or not those particles are gravitationally bound.  The excess of low mass halos indicates that FOF is finding spurious objects based on chance close groupings of particles, and that these groups cannot be confidently classified as gravitationally bound halos.  Standard FOF halos finders are known to have this behavior \citep{2011MNRAS.415.2293K}.  We are prevented from using a more sophisticated halo finder by the nature of our simulations:  particle-mesh methods suppress halo potential due the finite resolution of the simulation \citep{1988csup.book.....H}.  The minimum dark matter halo mass capable of hosting a Population III protostar is $1.5\times10^5~$M$_\odot$ \citep{2007ApJ...654...66O}, which is below the minimum halo mass that can be resolved in the $7.0~h^{-1}$Mpc boxes.  In an effort to include as many applicable halos as possible in the $7.0~h^{-1}$Mpc boxes, we use all halos identified by the halo finder, the smallest of which are composed of 8 particles, corresponding to a halo mass of $2.29\times10^5~$M$_\odot$.  There is a possibility that some of the smallest groups identified by the halo finder are chance arrangements of particles that are not actually gravitationally bound, but due to the potential importance of low mass halos as locations of Population III star formation they are included in the halo catalogs.  A minimum halo mass of $1.0\times10^5~$M$_\odot$, corresponding to 28 particles, is applied to the $3.5~h^{-1}$Mpc boxes.  This choice increases the confidence that the selected halos were bound objects and constrained the analysis to halos that were in agreement with the low-mass end of the Warren mass function \citep{2006ApJ...646..881W}.  At high halo masses, the mass functions for the simulations are suppressed due to the finite size of the simulation volume.  The $7.0~h^{-1}$Mpc boxes somewhat underproduce low-mass halos at early times due to the finite spatial resolution of the gravitational force; the $3.5~h^{-1}$Mpc boxes do the same, but at low redshift due to their small volume.  Overall, we are capturing the halo mass functions within a factor of 2 of the analytic results in the mass range and redshifts of interest, giving us confidence that the foundation upon which our semi-analytic model rests is solid and that our cosmological volumes of choice are reasonable.

\section{Model Description}
\label{sec:model}

\subsection{Overview}
The primary goal of our model is to determine the ability of a halo to form primordial stars.  To accomplish this, we start at the highest redshift in a simulation where dark matter halos are found, and then move forward in time, traversing our merger trees to follow the metal enrichment history of each halo over time, determining which halos are capable of forming a Population III star and when that formation ought to occur.  If a halo can form a Population III star, we assume that it does, and that the gas in that halo will in turn be enriched with metals.  Moving forward in time, halos are determined to be either chemically pristine or chemically enriched.  A halo is considered chemically pristine if none of its previous generations have hosted a star, while a halo with one or more parents that have hosted a star is considered to be chemically enriched.  Under the assumption of rapid mixing, any chemically enriched halo is by definition incapable of forming a Population III star, but can form chemically enriched stars.  The determination of chemically pristine and chemically enriched halos occurs at every simulation output, while numerical integration forward in time between two consecutive data outputs allows for more temporally-refined modeling of star formation, feedback, and ejection of material from the halos.  The halo catalogs and masses are used to establish the boundary conditions for the numerical integration of chemically enriched star formation, as detailed in \S\ref{sec:chemically enriched sf}.  This allows for much higher temporal resolution of the development of the stellar population, tracking of stellar ages, chemical enrichment, and gas ejection from each halo. 

A summary of the parameters of the model that were tested, their fiducial values, and the ranges investigated is given in Table \ref{table:parameters}.

\begin{table*}[htdp]
\caption{Model parameters with their fiducial value, the range tested, and a brief description.}
\begin{center}
\begin{tabular}{ l c c l }
\multicolumn{4}{c}{Model Parameters}	\\
\hline
\hline
Parameter	&	Fiducial Value	&	Range	&	Description	\\
\hline
$\varepsilon$	&	0.04		&	0.008-0.2					&	Star formation efficiency \\
$f_{\rm{esc}}^{\rm{LW}}$	&	1		&	0.01, 0.1, 1					&	LW photon escape fraction	\\
IMF				&	Salpeter	&	Salpeter, Kroupa, Chabrier	&	Chemically enriched stellar IMF	\\
$\varepsilon_{\rm{SN}}$	&	0.0015	&	0.0003-0.0075		&	SNe energy coupling	\\
\hline
\end{tabular}
\end{center}
\label{table:parameters}
\end{table*}

\subsection{Population III Star Formation}
\label{sec:pop III sf}
Population III stars are metal-free, and hence their formation hinges on a halo cooling via H$_2$ to a temperature and density at which its core is gravitationally unstable and collapses \citep{2002Sci...295...93A,2007ApJ...654...66O}.  The model we adopted for Population III star formation is based on the model of \citet{2009ApJ...694..879T}, and as in their work we determine the minimum mass of a halo capable of forming a Population III star by determining the mass that is required in order to cool to its virial temperature via H$_2$ cooling in the local Hubble time in the absence of H$_2$ photodissociating radiation.  Following \citet{2009ApJ...694..879T}, the cooling time is calculated as 
\begin{equation}
\tau_{\rm{cool}} = \frac{3k_B T_{\rm{vir}}}{2\Gamma f_{\rm{H}_2}},
\label{t_cool}
\end{equation}
where $k_B$ is Boltzmann's constant, $T_{\rm{vir}}$ is the halo virial temperature, $\Gamma$ is the H$_2$ cooling function, and $f_{\rm{H}_2}$ is the ratio of molecular to atomic hydrogen.  We adopt the H$_2$ cooling function of \citet{1998A&A...335..403G} as adapted by \citet{2009ApJ...694..879T} of
\begin{equation}
\Gamma \simeq 10^{-31.6} \left(\frac{T}{100~\rm{K}} \right)^{3.4} \left( \frac{n_{\rm{H}}}{10^{-4} \rm{cm}^{-3}} \right) ~ \rm{erg~s}^{-1},
\label{cooling_function}
\end{equation}
where $n_{\rm{H}}$ is the hydrogen number density.  Since the cooling time is long in a chemically pristine halo, the time lag between gas reaching the virial density and star formation occurring is very high, typically on the order of several tens of million years \citep{2007ApJ...654...66O}.  To make the lag less severe we assume that star formation occurs at a density that is higher than the virial density.  Reflecting this, our model assumes that the hydrogen number density in a halo at the time of collapse will be 10 times the virial density, rather than the virial density itself, as used by \citet{2009ApJ...694..879T}.  By taking the temperature in equation \eqref{cooling_function} to be the virial temperature, estimating $n_{\rm{H}}$ as 10 times the virial density, substituting these quantities into equation \eqref{t_cool} and equating this to the local Hubble time, the minimum H$_2$ fraction required for the halo to cool within the local Hubble time can be calculated.  This can in turn be equated to the maximum H$_2$ fraction that can form in a halo, as found by \citet{1997ApJ...474....1T},
\begin{equation}
f_{\rm{H}_2} \simeq 3.5\times10^{-4}\left(\frac{T}{1000\rm{K}}\right)^{1.52},
\label{max_fh2}
\end{equation}
and upon taking the temperature in equation \eqref{max_fh2} to be the virial temperature, the minimum halo mass capable of cooling via H$_2$ as a function of redshift can be determined.  This process yields a minimum halo mass of 
\begin{equation}
M_{\rm{min,Hubble}}=5.87\times 10^4 \left(\frac{1+z}{31}\right)^{-2.074}\rm{M}_\odot .
\label{m_min_hubble}
\end{equation}
The modification to the hydrogen number density in the halo is the only departure from the work of \citet{2009ApJ...694..879T}, and results in a reduction of the minimum halo mass required to cool within a Hubble time by a factor of approximately $2.6$.  This is more in agreement with recent simulations (e.g., \citet{2012ApJ...745...50W,2012MNRAS.427..311W}) where star formation occurs at lower masses than one would estimate using the values of \citet{2009ApJ...694..879T}.

The presence of other stars will modify this limit with the introduction of radiation capable of photodissociating H$_2$.  This radiation must be accounted for to accurately determine which halos are capable of forming a Population III star.  Radiation in the Lyman-Werner (LW) band ($11.18-13.60$ eV) can photodissociate H$_2$, suppressing cooling in a halo and requiring a larger halo mass in order to collapse \citep{2001ApJ...548..509M,2008ApJ...673...14O}.  Lyman-Werner radiation can similarly dissociate the HD molecule \citep{2011MNRAS.412.2603W}, but HD is the dominant coolant only below a temperature of approximately $150~$K, and cooling via H$_2$ is required to approach this regime.  To enable the formation of H$_2$ and effective cooling of metal free gas, the H$_2$ formation timescale must be at least equal the H$_2$ destruction timescale, as set by the rate of photodissociation due to LW radiation.  The formation timescale can be written as a function of the H$_2$ fraction, and the dissociation timescale can be written as a function of the proper LW flux.  Equating these timescales, solving for the H$_2$ fraction, and using this to evaluate the halo mass needed to cool in in the local Hubble time gives the minimum mass for a halo to cool efficiently via H$_2$ in the presence of a photodissociating background as 
\begin{equation}
M_{\rm{min,LW}}=1.91\times 10^6 J_{21}^{0.457}\left(\frac{1+z}{31}\right)^{-2.186}\rm{M}_\odot,
\label{m_min_LW}
\end{equation}
where $J_{21}$ is the proper LW flux.  Equation \ref{m_min_LW} is the case in which there is no H$_2$ self-shielding in the halo.  This limit is modified for each halo to account for self-shielding, and is addressed in \S\ref{sec:lyman_werner}.  $J_{21}$ is defined from the comoving LW  photon number density, $n_{\rm{LW}}$, in Mpc$^{-3}$, as
\begin{equation}
J_{21}=1.6\times10^{-65}n_{\rm{LW}}\left(\frac{1+z}{31}\right)^3~\rm{erg~s^{-1}~cm^{-2}~Hz^{-1}~sr^{-1}}.
\label{j21}
\end{equation}
The method for the determination of the LW flux is addressed in \S\ref{sec:lyman_werner}.  The minimum halo mass required for collapse and Population III star formation at a given redshift and LW flux will then be the maximum of equations \eqref{m_min_hubble} and \eqref{m_min_LW},
\begin{equation}
M_{\rm{min}}=\rm{max}
	\begin{cases}
	5.87\times 10^4 \left(\frac{1+z}{31}\right)^{-2.074}\rm{M}_\odot	\\
	1.91\times 10^6 J_{21}^{0.457}\left(\frac{1+z}{31}\right)^{-2.186}\rm{M}_\odot
	\end{cases}.
	\label{minmass}
\end{equation}
The reader is encouraged to see \citet{2009ApJ...694..879T} for details on the derivation of the minimum halo mass requirements.

When a chemically pristine halo with a mass sufficiently large for Population III star formation is identified, our model assumes that a star forms.  The halo is given a mass of gas equal to its dark matter mass multiplied by $\Omega_B/\Omega_{DM}$.  This halo is then tagged as being chemically enriched, and its child halos are no longer capable of forming Population III stars at later times.

Population III stars are assumed to end their lives as Type II supernovae (SNII).  This explosion can expel gas from the halo and delay the start of chemically enriched star formation.  To account for this delay, after a Population III star forms, the host halo is tagged with a delay time equal to the sum of the assumed Population III stellar lifetime and a delay time of 30 Myr.  During this time no chemically enriched star formation can occur in this halo if the halo either does not grow in mass or grows only by merging with other chemically pristine halos.  If the halo merges with a halo that is already hosting chemically enriched star formation, star formation in the combined halo is not stopped.

\subsection{Chemically Enriched Star Formation}
\label{sec:chemically enriched sf}
Halos containing particles that have previously been in a halo that formed stars are considered to be chemically enriched, and their star formation is treated differently than the chemically pristine halos that form Population III stars.  Chemically enriched halos inherit gas from the halos that merge to form them, with the difference between the sum of the mass of the merging halos and the current total halo mass treated as accreted pristine material, contributing a mass of gas in the same manner as gas is added to a pristine halo.  Star formation in chemically enriched halos is assumed to be continuous and a function of the mass of gas available in the halo.  The rate of growth of stellar mass in a halo is taken to be a function of the mass of gas in that halo \citep{2010ApJ...724..687L} and is modeled as 
\begin{equation} 
\frac{dM_\star}{dt} = \frac{\varepsilon}{\tau}M_{\rm{gas}}(t),
\label{dmstardt}
\end{equation}
where $\varepsilon$ is the dimensionless star formation efficiency and $\tau$ is the characteristic star formation time, taken here to be $10^8$ yr.  The quotient $\varepsilon/\tau$ with the fiducial star formation efficiency of this work, $\varepsilon=0.04$, is similar to the galactic gas depletion time of a few Gyr \citep{2011ApJ...730L..13B}.  Equation \eqref{dmstardt} is integrated forward in time across simulation outputs using a fourth order Runge Kutta method.  The integration timestep is one percent of the elapsed time between the current and subsequent data outputs, so each data output is traversed in 100 integration steps, with results being insensitive to the precise choice of timestep size.  At every integration time step, stars are formed and the gas reservoir in the halo is decremented by the same amount as the mass of stars created.  Tracking the star formation in enriched halos is achieved by following the total stellar mass formed rather than individual stars, as this removes assumptions about the initial mass function (IMF) from the star formation process (though the chemical feedback depends on on the IMF; see \S\ref{sec:lyman_werner} and \ref{sec:halo_gas_ejection}).  The chemical feedback from the stellar population in a halo back to the halo gas is a function of the age of the stellar population.  Stars are created at every integration time step, so utilizing this age-dependent feedback model requires that the age distribution of the stellar population in each halo is tracked.  The entire stellar population is followed across 100 age bins at each data output, with the bins equally spaced from the time that the first star in the simulation formed to the time of the next data output.  The time of the next output corresponds to the end boundary condition of the time integration of equation \eqref{dmstardt} in the current data output.  At each integration timestep, the stellar mass in each age bin returns material to the interstellar medium (ISM) equal to the product of the stellar mass in that age bin, the integration timestep, and the mass of gas and metals that are ejected per solar mass of stars per year from the tabulated values that correspond to that stellar age and metallicity.  As equation \eqref{dmstardt} is integrated forward in time, the mass in the stellar age bins are advanced forward to accurately reflect the age distribution at any given time.  When the model advances to a new data output, the chemical, gas, and stellar content, complete with age distribution of each parent halo's stellar population is inherited by the child halo in proportion with the fraction of the parent halo mass that was passed to the child halo.  At this point the stellar age bins are reconstructed to reflect the new, longer, timespan imposed by the end time of the new dataset, and the stellar population is remapped to these new age bins.  This decreases the time resolution of the population ages, but maintains the character of the distribution, and is done to mitigate computational memory usage.  In practice, the largest age bin in the model spans $3.72$ Myr, which is larger than only the smallest stellar age bin in the tabulated data of material returned to the ISM by the stellar population, resulting in the temporal resolution of the feedback model being primarily limited to the temporal resolution of the available stellar feedback data.  A full description of the chemical evolution model and the results from it are presented in Paper II.

\subsection{Lyman Werner Flux Determination}
\label{sec:lyman_werner}
Determining the number density of Lyman-Werner (LW) photons produced by the stars in the simulation requires determining the star formation rate (SFR) for both Population III and chemically enriched stars.  The Population III SFR is determined by multiplying the number of chemically pristine halos massive enough to host a Population III star by a user-determined characteristic Population III stellar mass and dividing by the simulation volume and output time step.  This is further multiplied by a factor representing the Population III stellar multiplicity in each halo, giving the total mass of Population III stars formed per year per comoving Mpc$^3$ during the time spanned by this simulation output.  The characteristic Population III mass is taken in this work to have a fiducial value of $30$ M$_\odot$ \citep{2006ApJ...641....1T}.  The multiplicity factor is a user-defined parameter that was set to a fiducial value of $1.2$, drawing inspiration from the findings of \citet{2009Sci...325..601T} which observe fragmentation of the pre-stellar cloud, suggesting the possibility of the formation of Population III binary star systems.  A recent constraint on the Population III binary fraction of $36\%$ is reported by \citet{2012arXiv1211.1889S}.  This value was not used in our model, but changes to the multiplicity factor will modify the Population III SFR as a direct multiplicative factor, and does very little to change the total SFR as chemically enriched star formation generally dominates the Population III SFR by several orders of magnitude, as shown in \S\ref{sec:results}.  Similarly, the LW flux is dominated by chemically enriched stars, and the minimum halo mass for Population III star formation does not change significantly with changes to the assumed Population III multiplicity.  The Population III SFR is then used to determine the density of LW photons by multiplying by an estimated Population III stellar lifetime of $2.5$ Myr \citep{2003A&A...397..527S} and the rate of LW photon production per solar mass per year by a metal free star from \citet{2003A&A...397..527S}.

Determining the number density of LW photons from chemically enriched stars follows a similar process.  The total stellar mass in all chemically enriched halos is summed and multiplied by the rate of LW photon production per solar mass per year by a chemically enriched, continuously star forming population from \citet{2003A&A...397..527S}, which assumes a Salpeter initial mass function.  We do not modify this value for different initial mass functions.  

The LW photon production rates for both Population III and chemically enriched stars is modified by multiplying by the LW photon escape fraction, $f_{\rm{esc}}^{\rm{LW}}$, to approximate the effects of absorption within the halo in which they originate.  This is a user-defined parameter with a fiducial value of $1$ and other explored values of $0.01$ and $0.1$ and is independent of halo mass.  These values were adopted following \citet{2004ApJ...613..631K}, who investigate the LW photon escape fraction as a function of stellar mass, halo mass, and redshift (among other things) for an individual star in a halo, finding that the LW escape fraction decreases with increasing halo mass and decreasing stellar mass.  This is obviously not equivalent to the scenario of chemically enriched star formation in our model, but can be used as a guiding approximation by taking the total mass of stars in a halo to be akin to the individual stellar mass of \citet{2004ApJ...613..631K}.  The ratio of stellar mass to halo mass in our model exceeds the corresponding ratio required for an escape fraction of $0.01$ in \citet{2004ApJ...613..631K}, for the most stringent case of a $500~$M$_\odot$ star, in more than 98 percent of halos at $z=10$.  More than 97 percent of halos in our model meet the criteria for a LW escape fraction of $0.1$, and 88 percent meet the criteria for a LW escape fraction of 1.  While our comparison to \citet{2004ApJ...613..631K} makes numerous simplifying assumptions, it suggests that the effective LW escape fraction in our model is almost certainly greater than $0.01$, and is very likely greater than $0.1$.

The total LW photon density is taken as the sum of the LW densities from Population III and chemically enriched sources that have not been redshifted out of the LW band.  Using the middle of the LW band as an average, photons produced at redshift $z_{\rm{p}}$ will be redshifted out of this band at a redshift
\begin{equation}
z_{\rm{exit}}=(z_{\rm{p}}+1)\left(\frac{11.18}{12.39}\right)-1
\end{equation}
and will no longer able to dissociate H$_2$.  For example, LW photons produced at $z=20$ will be redshifted out of the LW band at $z=17.9$.  Once the cumulative LW photon density has been determined, the LW flux, $J_{21}$, can be computed as described in \S\ref{sec:pop III sf}.  This is done based on the stellar content in the simulation at a given time, and is done as the model progresses through the merger tree.  Self-shielding of H$_2$ to LW radiation can in principle have a significant impact on the ability of the halo to cool \citep{2011MNRAS.418..838W}.  H$_2$ self-shielding reduces the effective LW flux in a given halo, which in turn reduces the minimum halo mass necessary for Population III star formation in the presence of a photodissociating background.  A prescription for H$_2$ self-shielding following \citet{2011MNRAS.418..838W} is implemented to determine the shielding factor $f_{\rm{sh}}$ by which $J_{21}$ is reduced.  The shielding factor is calculated as 
\begin{equation}
f_{\rm{sh}}\left(N_{\rm{H}_2}, b\right) = \frac{0.956}{\left(1+x/b_5\right)^\alpha}+\frac{0.035}{\left(1+x\right)^{0.5}}\exp{\left[-8.5\times10^{-4}\left(1+x\right)^{0.5}\right]}~,
\end{equation}
\label{f_sh}
where $x$ is the scaled $H_2$ column density, $x=N_{\rm{H}_2}/5\times10^{14}~$cm$^{-2}$, $b$ is the scaled Doppler broadening parameter, $b_5=b/10^5~$cm s$^{-1}$, and the parameter $\alpha$ takes a value of $1.1$.  The effective $J_{21}$, defined as $J_{21,\rm{eff}}=f_{\rm{sh}}J_{21}$, is used in Equation \ref{m_min_LW} to determine whether that halo is sufficiently massive to form a Population III star.

\subsection{Halo Gas Ejection}
\label{sec:halo_gas_ejection}
The chemical evolution and feedback from pristine and chemically enriched halos are treated separately.  Population III stars are taken to eject a fixed mass of metals to the interstellar medium (ISM) at their death, set by yields of Type II supernova (SNII) calculations \citep{2002ApJ...567..532H}. Gas and metals are also ejected from the halo to the intergalactic medium (IGM) following a method similar to \citet{2010ApJ...708.1398T}, which compares the kinetic energy of the supernova driven wind in the halo to the escape velocity of the halo at the virial radius.  The gas mass ejected due to supernova-driven winds is calculated as 
\begin{equation}
M_{\rm{lost}}=3.9\times10^8\left(\frac{N_{\rm{SN}}\varepsilon_{\rm{SN}}E_{51}r_{\rm{vir}}}{GM_{\rm{vir}}}\right)M_\odot~,
\label{m_lost}
\end{equation}
where $N_{\rm{SN}}$ is the number of supernovae that occurred in the current integration timestep, $\varepsilon_{\rm{SN}}$ is the efficiency with which supernova energy is converted to the kinetic energy of the wind, $E_{51}$ is the supernova energy in units of $10^{51}~$erg, $r_{\rm{vir}}$ is the virial radius in units of proper Mpc, $G$ is the gravitational constant in CGS units, and $M_{\rm{vir}}$ is the virial mass in solar masses.  

$N_{\rm{SN}}$ is determined as the multiplicity factor multiplied by the number of Population III stars in a halo (always taken to be one in this work) for chemically pristine halos, and in chemically enriched halos it is found from tabulated supernova rates given the age of a stellar population and the adopted stellar initial mass function.  $E_{51}$ and $\varepsilon_{\rm{SN}}$ are user-defined parameters that can be varied.  $\varepsilon_{\rm{SN}}$ is given a fiducial value of $0.0015$, following \citet{2010ApJ...708.1398T}, which assumes that $5\%$ of the total supernova energy is kinetic, and that of this $3\%$ is transferred to the ejected material.  Changes to $\varepsilon_{\rm{SN}}$ have a negligible impact on the star formation rate (SFR) density, as the amount of gas ejected from a halo is very small compared to the reservoir of gas available for star formation in that halo.  The change in the total amount of gas in a halo that arises from variation of $\varepsilon_{\rm{SN}}$ is in turn negligibly small, though it has a significant impact on the chemical evolution of the halo, as discussed in Paper II.

\begin{figure}[htbp] 
   \centering
   \includegraphics[width=.45\textwidth, clip=true]{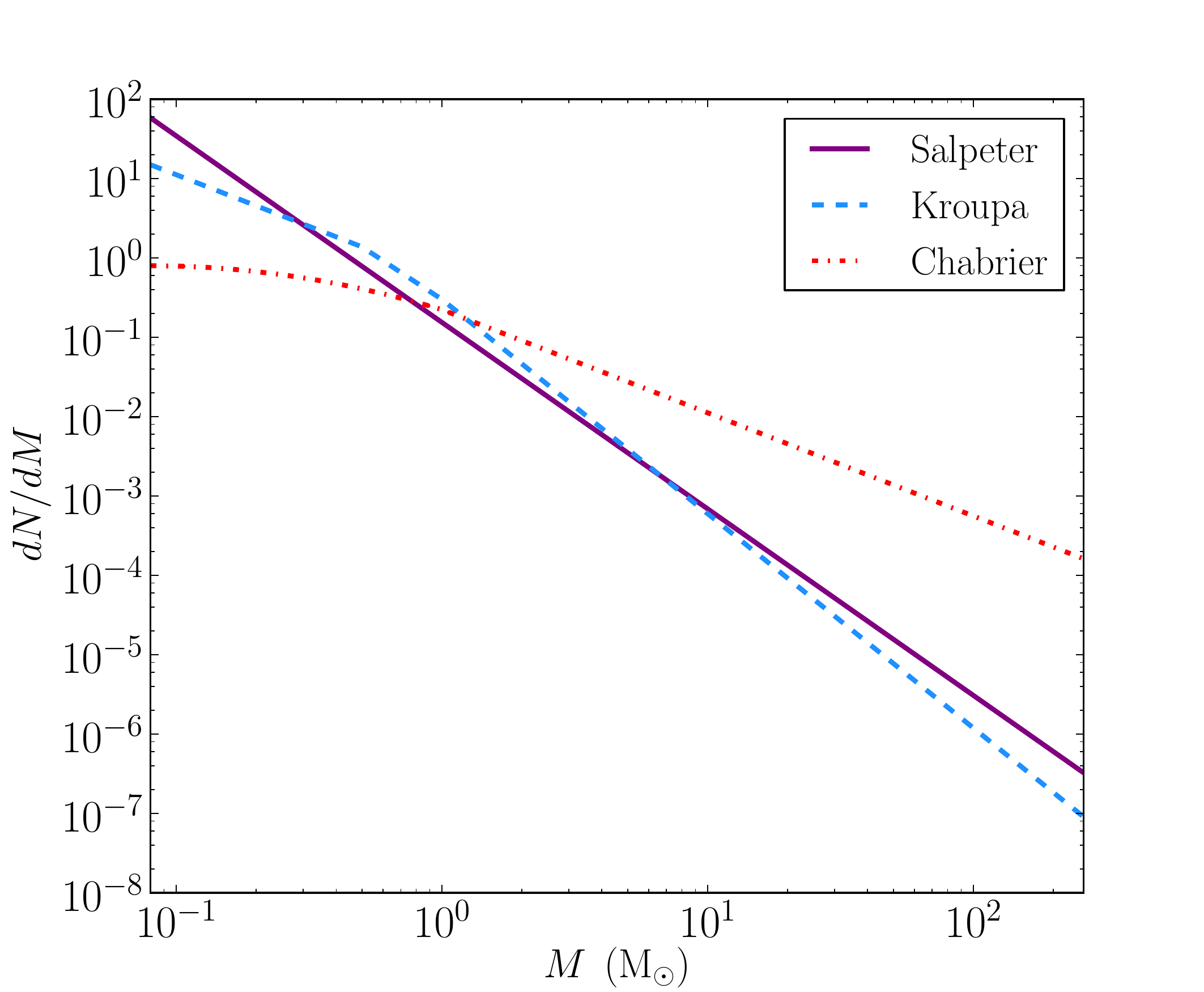} 
   \caption{The three IMFs considered in this work are Salpeter (violet, solid line), Kroupa (blue, dashed line), and Chabrier (red, dot-dashed line), over a mass range of $0.08M_\odot$ to $260M_\odot$.  The integrated area under each of the curves is the same.  The Salpeter IMF emphasizes low mass stars, the Kroupa IMF emphasizes intermediate mass stars, and the Chabrier IMF is by the far the most top-heavy of the three, emphasizing high mass stars.}
   \label{fig:imf_plot}
\end{figure}

Yields as a function of stellar age were convolved with an integrated initial mass function (IMF) to create a table of the total ejected gas mass (Peruta et al. 2013, submitted) in units of solar masses of metals per year per solar mass of stars.  Tables in this form were created for Salpeter, Chabrier, and Kroupa IMFs, at various metallicities, and separately for the Type Ia supernovae (SNIa) \citep{1999ApJS..125..439I,2009ApJ...707.1466K} and the combined ejecta of SNII and asymptotic giant branch (AGB) stars \citep{2010MNRAS.403.1413K,2006NuPhA.777..424N}.  Two sets of these tables were created, one for stars at and above solar metallicity, and one for stars below solar metallicity.  This process was repeated for each of the chemical species that is tracked.  While changes to the IMF primarily impact the chemical evolution of the stellar populations, it influences the bulk star formation properties by changing the amount of gas returned to the ISM that is available for star formation.  The three IMFs have functional forms
\begin{eqnarray}
\frac{dN}{dm}=&\Phi_{\rm{Salpeter}}=&0.154m^{-2.35}\\
&\Phi_{\rm{Kroupa}}=&\left\{ \begin{array}{lc}
0.56m^{-1.3}& m\le0.5M_{\odot}\\ 
0.3m^{-2.2}&0.5M_{\odot} < m \le1M_{\odot}\\ 
0.3m^{-2.7}& m> 1M_{\odot}
\end{array}\right.\\
&\Phi_{\rm{Chabrier}}=& \left\{\begin{array}{lc} 
0.799 e^{-(\log m/m_{\rm c})^2/2\sigma^2}&  m \leq 1 M_{\odot} \\ 
0.223 m^{-1.3} & m > 1 M_{\odot}
\end{array}\right.\\
\end{eqnarray}
where in the Chabrier IMF $m_c$ is the characteristic mass and takes a value of $0.079M_\odot$ and the dispersion $\sigma=0.69$ \citep{1955ApJ...121..161S,2002Sci...295...82K,2003PASP..115..763C}.  The IMFs were considered over a range of masses from $0.08M_\odot$ to $260M_\odot$, and are shown in Figure \ref{fig:imf_plot}.  As equation \eqref{dmstardt} is integrated forward in time for each halo, the stellar mass in each stellar age bin is multiplied by the integration time step and the normalized yield corresponding to the metallicity and age of the stellar population to determine the mass of metals ejected to the ISM via SNII and AGB stars.  The total mass ejected from a given stellar age bin is removed from the mass of stars in that bin and returned to the ISM.

Tables similar those giving the chemical yields and gas mass ejected from stars were created giving the rate of SNIa, in units of number of SNIa per year per solar mass of stars, and are used to calculate the total number of SNIa expected to occur in the stellar population of a given halo in an integration time step.  This is used with the mass of ejecta from SNIa and equation \eqref{m_lost} to determine the mass of gas ejected to the IGM from the halo as a result of SNIa explosions.  The interested reader is directed to Peruta et al. (2013; submitted) for more information.

The gas ejection model of equation \eqref{m_lost} would potentially allow for all of the gas to be ejected from a halo following the death of a Population III star.  The model has the capability to treat these halos as chemically pristine, as they do not have any metals in the ISM, and to allow them to merge with other pristine halos and form Population III stars.  In practice this almost never happens, and any halo that forms stars retains some chemically enriched material, rendering it incapable of forming a Population III star.

\section{Results}
\label{sec:results}
\subsection{Overview}
This project endeavors to develop a semi-analytic model that can be used in conjunction with a cosmological simulation to create a catalog of halos capable of forming Population III stars.  The halos in this catalog extend beyond the first, most massive object in the simulation to encompass chemically pristine halos across a wide range of redshifts and environments while self-consistently accounting for the halo merger history and the photodissociating radiation produced by the stars in the simulation.  In this paper the initial mass function (IMF) of chemically enriched stars was varied between three commonly used IMFs: Salpeter, Kroupa, and Chabrier.  An additional free parameter that is varied is the star formation efficiency, $\varepsilon$, which adopts a fiducial value of $0.04$, following \citet{2005ApJ...629..615W}.  The star formation efficiency is varied from this fiducial value by a factor of 5 in both directions, investigating values similar to those explored by \citet{2009ApJ...694..879T}.  This range also encompasses the equivalent star formation efficiency of $0.01$ adopted as the fiducial value by \citet{2010ApJ...708.1398T}.  The Lyman-Werner photon escape fraction, $f_{\rm{esc}}^{\rm{LW}}$, was varied from its fiducial value of 1 to $0.1$ and $0.01$.

\begin{figure}[htbp] 
   \centering
   \includegraphics[width=.45\textwidth, clip=true]{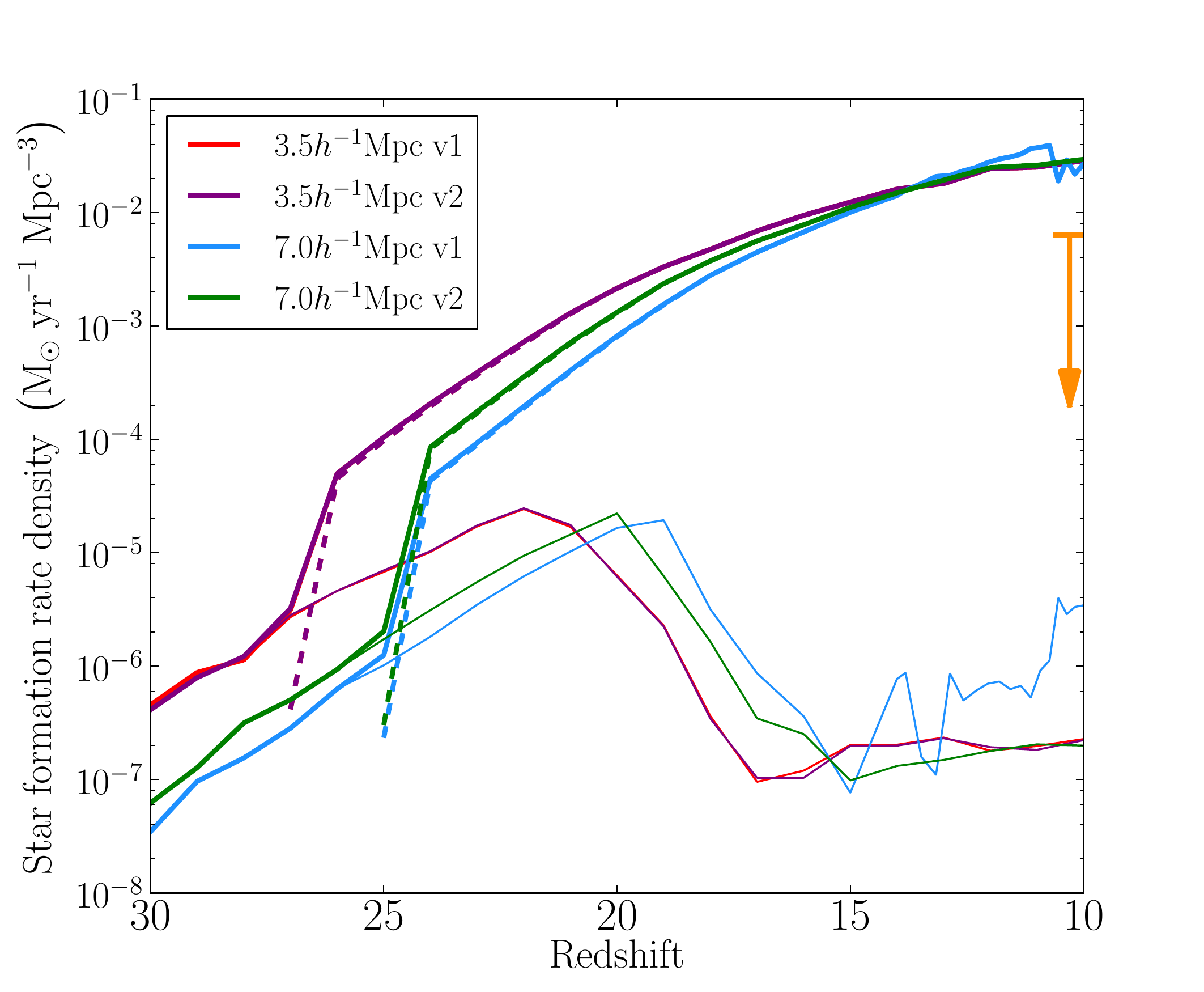} 
   \caption{The star formation rate (SFR) densities for all four simulations for our fiducial choice of parameters.  The thick, solid line shows the total SFR density, the dashed line is the chemically enriched SFR density, and the thin, solid line is the Population III SFR density.  An extrapolated observational upper limit from \citet{2011ApJ...737...90B} is shown in orange.  All four simulations show very good agreement despite testing two different volumes and all being created from initial conditions generated with different random seeds.}
   \label{fig:sfr_all_boxes}
\end{figure}

To assess the validity of this model, comparison is made between the predicted star formation rate (SFR) density and observational limits.  The SFR density is primarily sensitive to the star formation efficiency, $\varepsilon$.  Figure \ref{fig:sfr_all_boxes} shows the comoving SFR densities of all four of the simulations in units of M$_\odot$yr$^{-1}$Mpc$^{-3}$.  All of the simulations show very similar total, Population III, and chemically enriched SFR densities.  All simulations show the same qualitative behavior, and the variations in the onset of star formation and the spread in SFR densities at $z=10$ are all within a range that would be expected from variations due to differing large scale structure (i.e., cosmic variance) in different simulations.  In particular, at late times all halos converge to very similar values as the total number of halos increases and stochasticity, as a result, decreases.  This convergence lends support to the notion that our simulation volumes are all effectively statistically equivalent to each other.  As a result, and for clarity, for the remainder of this paper all discussion and the content of all plots will be limited to one $7.0~h^{-1}$Mpc box (labeled $7.0~h^{-1}$Mpc v1) unless otherwise noted, as in \S\ref{sec:5Mpc_results}.  Using this simulation, the model produces a catalog of more than 40,000 halos that are capable of forming a Population III star.  These halos span a redshift range from $z=30$ to $z=10$, and range in mass from $2.3\times10^5~$M$_\odot$ to $1.2\times10^{10}~$M$_\odot$.

\subsection{Star Formation Rates}
\begin{figure*}[htbp] 
   \centering
   \includegraphics[width=.9\textwidth, clip=true]{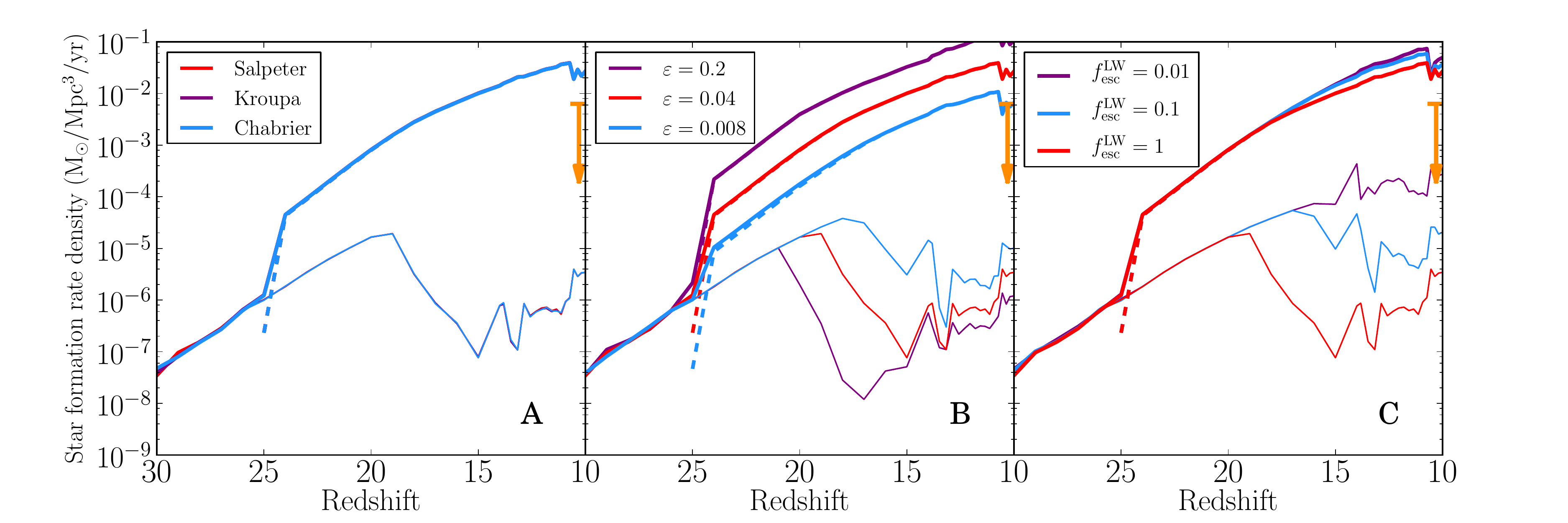} 
   \caption{The star formation rate (SFR) density in M$_\odot$yr$^{-1}$Mpc$^{-3}$ as a function of redshift for variations in IMF, star formation efficiency, and Lyman-Werner (LW) photon escape fraction.  Panel A shows the effect of varying the IMF, Panel B shows the effect of varying the star formation efficiency, and Panel C shows the effect of varying the LW photon escape fraction.  In all panels, the Population III SFR density is plotted in thin, solid lines, the chemically enriched SFR density is plotted in dashed lines, and the total SFR density is plotted in thick, solid lines.  An extrapolated observational upper limit from \citet{2011ApJ...737...90B} is shown in orange.  In Panel A, the total and component SFR densities for the three IMFs are nearly indistinguishable.  Panel B shows the SFR density using a Salpeter IMF and varying the star formation efficiency.  Increasing the star formation efficiency increases the difference between the Population III and chemically enriched SFR densities, driving Population III star formation down and chemically enriched star formation up.  In Panel C, changes to the LW photon escape fraction have a small effect on the chemically enriched SFR density, and decreasing the escape fraction increases the Population III SFR density at late times.  In all cases, chemically enriched star formation rapidly dominates Population III star formation by several orders of magnitude, but Population III star formation continues at very low levels for the entirety of the simulation.}
   \label{fig:sfr_grid}
\end{figure*}

The star formation rate (SFR) density, defined as the SFR per comoving Mpc$^{3}$, was tracked individually for the Population III and chemically enriched stellar populations.  The SFR density was sensitive primarily to the star formation efficiency for metal-enriched stars in equation \eqref{dmstardt}, with larger values of the efficiency increasing the chemically enriched SFR density, and in turn driving up the total SFR density.  The SFR density for several different sets of parameters is shown in Figure \ref{fig:sfr_grid}.  While chemically enriched stars dominate the star formation at essentially all redshifts, Population III stars continue to form throughout the duration of the simulation, at roughly constant but vastly subdominant levels.  The transition from the total SFR density being dominated by Population III stars to being dominated by chemically enriched stars happens very rapidly, in less than 10 million years after the first stars in the simulation form.  By $z=10$, the Population III SFR density is 3-5 orders of magnitude lower than the chemically enriched SFR density.  The total SFR density at $z=10$ ranges from $6.4\times10^{-3}~$M$_\odot$yr$^{-1}$Mpc$^{-3}$, using a star formation efficiency of $8\times10^{-3}$, to $2.1\times10^{-1}~$M$_\odot$yr$^{-1}$Mpc$^{-3}$ using a star formation efficiency of $0.2$.  The same star formation efficiencies yield Population III SFR densities of $5.2\times10^{-5}$ and $1.2\times10^{-6}~$M$_\odot$yr$^{-1}$Mpc$^{-3}$, respectively.  Increasing the chemically enriched star formation efficiency suppresses Population III star formation by increasing the Lyman-Werner flux, driving up the minimum halo mass for Population III star formation.  Including feedback from the stellar population impacts the chemically enriched SFR density in two ways.  Gas returned to the interstellar medium (ISM) from stars increases the reservoir of gas available for star formation, but gas ejected from the halo to the intergalactic medium by supernovae reduces the available gas.  The net change in the gas content of the ISM as a result of these two effects will determine whether stellar feedback increases or decreases the SFR density at any given integration timestep.  The initial mass function (IMF) has no significant impact on the SFR density for Population III or chemically enriched stars.  The LW photon production is determined based on the total mass of stars that are formed, and is agnostic of the distribution of stellar masses.  The slight variations that between the SFR density with different IMFs arise from the differences in the mass of gas returned to the halo that is available for star formation and differences in the amount that is ejected from the halo via supernovae, both of which are very small when compared to the total reservoir of gas available for star formation in a given halo.  Decreasing the LW photon escape fraction increases the Population III SFR density at late times by suppressing the photodissociating background radiation, decreasing the mass threshold for Population III star formation in a halo.  The Population III star formation mass threshold depends on on the proper LW flux as $M_{\rm{threshold}} \sim J_{21}^{0.457}$, so the magnitude of the increase in the Population III SFR density closely follows the changes in the LW flux, as would be expected from a collection of chemically pristine halos with masses below the Population III star formation mass threshold.  The reduction in the chemically enriched SFR density that accompanies the increased LW photon escape fraction originates from the decrease in number of halos that are capable of forming Population III stars.  By not forming stars these halos are not chemically enriched, and are unable to become sites of future chemically enriched star formation.

\begin{figure*}[htbp] 
   \centering
   \includegraphics[width=.9\textwidth, clip=true]{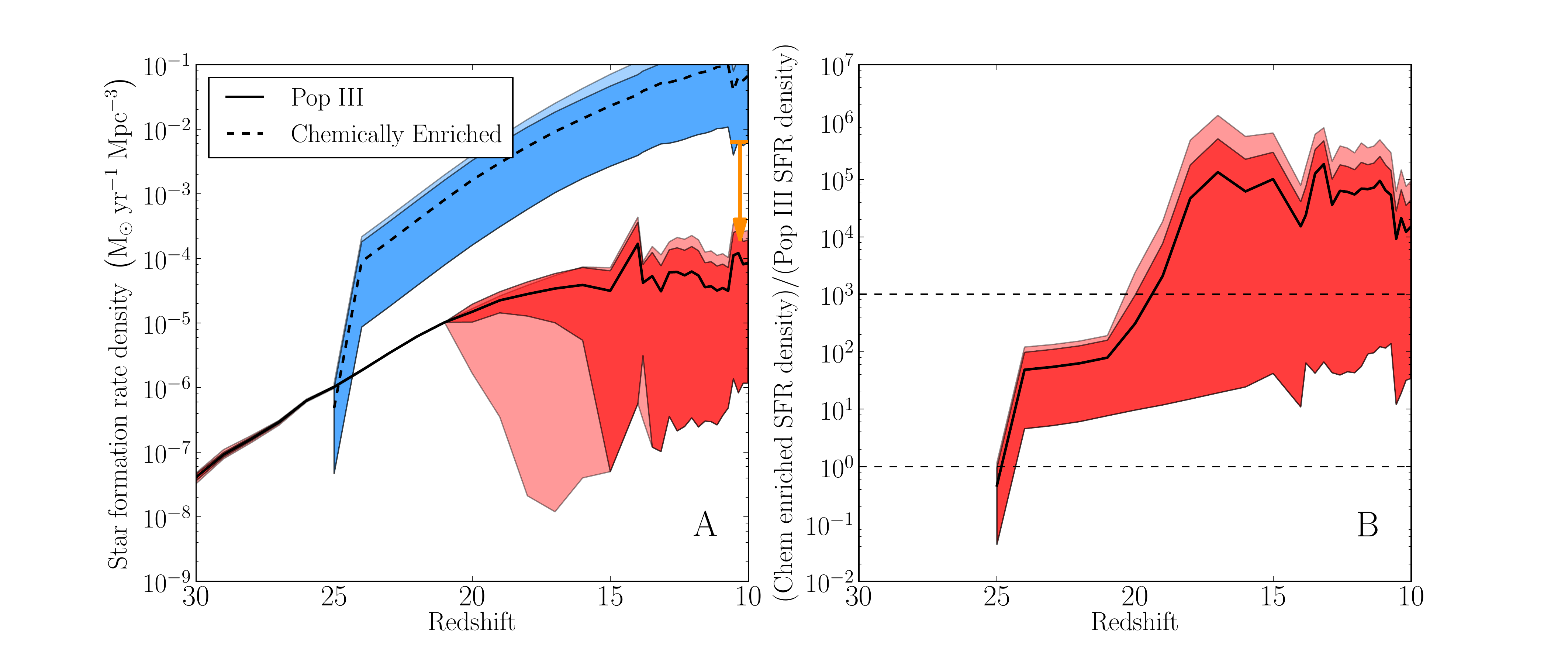} 
   \caption{Panel A shows the variation in star formation rate (SFR) density in M$_\odot$yr$^{-1}$Mpc$^{-3}$ as a function of redshift for all combinations of parameters in our model.  The mean Population III SFR density is plotted as a black solid line, and the average chemically enriched SFR density is plotted as a black dashed line.  The maximum range spanned by the Population III and chemically enriched SFR densities are shown by the light red and light blue shaded regions, respectively.  The dark shaded regions show the 68 percent confidence intervals around the mean.  An extrapolated observational upper limit from \citet{2011ApJ...737...90B} is shown in orange.  Panel B shows the ratio of the chemically enriched SFR density to the Population III SFR density as a function of redshift for all parameter combinations, with the black line showing the average value and the shaded regions having the same meaning as in Panel A.  To aid in interpretation, dashed lines are shown at levels corresponding to chemically enriched to Population III star formation rate ratios of 1 and 1000.}
   \label{fig:sfr_range}
\end{figure*}

Panel A of Figure \ref{fig:sfr_range} shows the ranges spanned by the Population III and chemically enriched SFR densities for all possible combinations of model parameters, along with the mean values and 68 percent confidence intervals.  The star formation efficiency is the parameter which creates the most variation in the chemically enriched SFR density.  There is no combination of parameters that can change the chemically enriched SFR density as greatly as the star formation efficiency, resulting in three distinct groupings in the SFR density corresponding to the three values of the star formation efficiency.  This is turn drives the confidence interval to encompass the majority of the range spanned by the maximum and minimum chemically enriched SFR densities.  The Population III SFR density is significantly effected by all parameters with the exception of the supernova efficiency, resulting in a confidence interval that is much smaller than the maximum and minimum SFR densities at a given redshift.  Panel B of Figure \ref{fig:sfr_range} shows the ratio of the chemically enriched SFR density to the Population III SFR density as a function of redshift.  The chemically enriched SFR density rapidly surpasses the Population III SFR density.  By $z=10$ the chemically enriched SFR density is a minimum of $1.9\times10^{3}$ times greater than the Population III SFR density, and is an average of $4.1\times10^{5}$ times greater than the Population III SFR density.

\subsection{Population III Star Formation Halo Mass Limit}
\begin{figure*}[htbp] 
   \centering
   \includegraphics[width=.9\textwidth, clip=true]{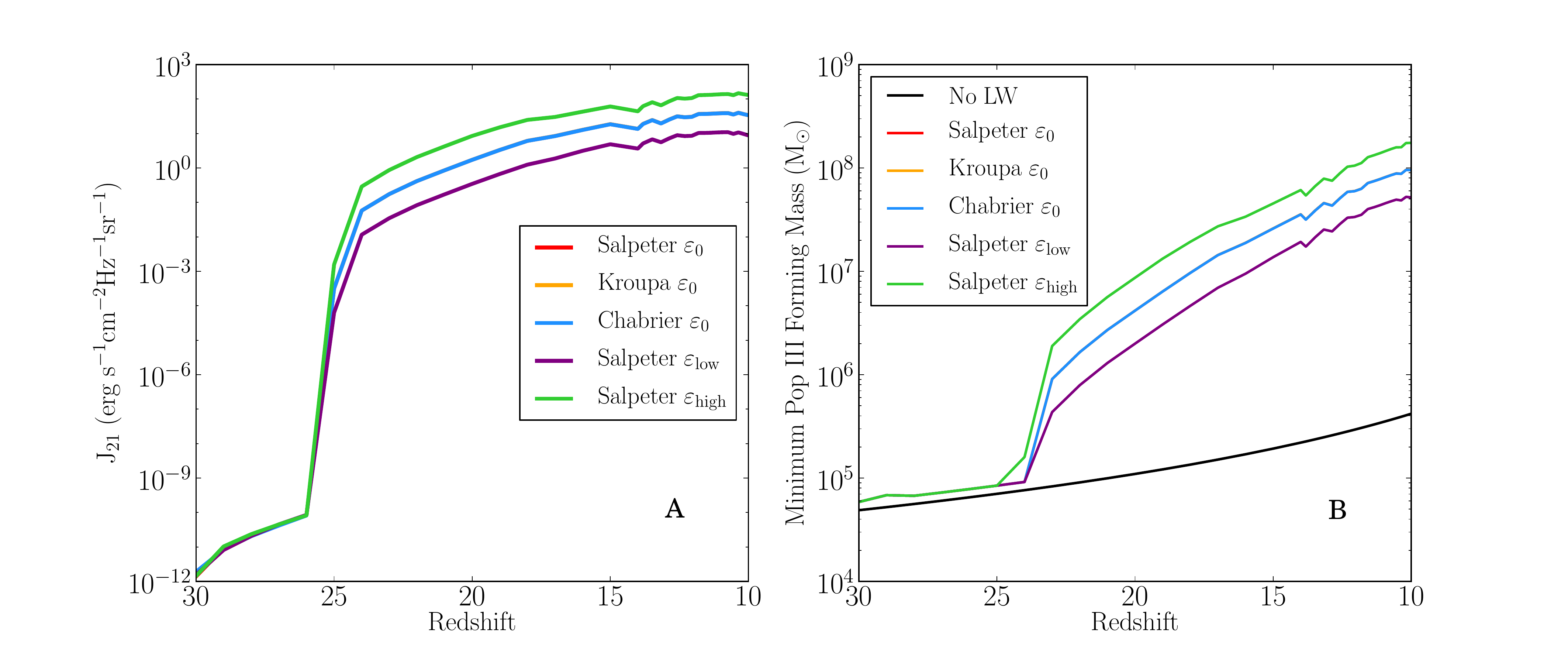} 
   \caption{The Panel A shows the proper Lyman-Werner flux ($J_{21}$) and Panel B shows the minimum mass threshold for Population III star formation as a function of redshift, plotted for the three initial mass functions (IMF) with a star formation efficiency of $0.04$ as well as for a Salpeter IMF with star formation efficiencies of $0.008$ and $0.2$.  The Salpeter, Kroupa, and Chabrier IMFs are indistinguishable in this plot.  For comparison, in Panel B the mass threshold without accounting for radiative feedback is shown in black, and is much lower, particularly once chemically enriched star formation become the dominant component of the stellar mass in the simulation.  The minimum halo mass for Population III star formation is dependent on $J_{21}$, and as a result Panel A closely mirrors the behavior of Panel B.  Simultaneously, $J_{21}$ is dependent on the stellar mass in the simulation, and will reflect the trends of the total star formation rate density in Figure \ref{fig:sfr_grid}.}
   \label{fig:j21_and_m_threshold}
\end{figure*}

The impact of the photodissociating radiation produced by the stars in the simulation on the minimum halo mass required for Population III stars formation is shown in Figure \ref{fig:j21_and_m_threshold}.  Panel A of Figure \ref{fig:j21_and_m_threshold} shows the proper LW flux, and Panel B plots equation \eqref{m_min_hubble}, giving the minimum halo mass required for Population III star formation in the absence of photodissociating radiation, along with equation \eqref{minmass}, which gives the minimum halo mass for Population III star formation when the photodissociating radiation produced by the stars in the simulation is accounted for.  At roughly $z=25$, the threshold mass accounting for the photodissociating radiation diverges from the threshold determined while neglecting the photodissociating radiation, as chemically enriched star formation begins to dominate the total star formation rate (SFR) density.  By the end of the simulation, at $z=10$, the threshold halo mass for Population III star formation has increased by two orders of magnitude due to photodissociating radiation from the stars in the simulation, drastically changing the membership and evolution of the set of halos capable of forming a Population III star throughout the simulation.  For reference, approximating the threshold masses at the beginning and end of the simulation shows that at $z=25$, a halo of mass $10^5~$M$_\odot$ has a virial temperature of approximately 260 K, and at $z=10$ a halo of mass $10^{7.5}~$M$_\odot$ will have a virial temperature of approximately 5190 K.

Figures \ref{fig:sfr_grid} and \ref{fig:j21_and_m_threshold} illustrate the relationship between the SFR density, the minimum halo mass for Population III star formation, and the H$_2$-photodissociating Lyman Werner (LW) flux.  The LW flux depends entirely on the stellar mass in the simulation volume and reflects the integrated SFR density.  The minimum halo mass for Population III formation is a function of redshift and the LW flux, and as more H$_2$ is photodissociated, increasingly larger halos are required in order to form Population III stars.  This in turn drives down the Population III SFR density but does not hinder chemically enriched star formation where, due to their metal content, despite the LW flux halos can still cool efficiently and form stars as H$_2$ is no longer the primary coolant.  Increasing the star formation efficiency will therefore increase the chemically enriched SFR density, which increases the LW flux, driving up the minimum halo mass for Population III formation, and in turn driving the Population III SFR density down.  Increasing the star formation efficiency not only increases the minimum halo mass for Population III star formation, but also changes the environment of these halos as times goes by, as will be described in \S\ref{sec:halo_environment}.  This cycle is responsible for the divergence of the Population III and chemically enriched SFR densities as the star formation efficiency is increased.

\subsection{$3.5~h^{-1}$Mpc Box Star Formation Rates}
\label{sec:5Mpc_results}
The $3.5~h^{-1}$Mpc simulations provide an opportunity to investigate the impact of low mass halos on the star formation rate (SFR) density.  These simulations have 8 times better mass resolution than the $7.0~h^{-1}$Mpc simulations, and the smallest halo objects identified by the Friends-of-Friends \citep{1985ApJS...57..241E} halo finder have a minimum mass of $2.86\times10^4~$M$_\odot$, significantly below the minimum halo mass of $1.5\times10^5~$M$_\odot$ \citep{2007ApJ...654...66O} required to host Population III star formation.

\begin{figure*}[htbp] 
   \centering
   \includegraphics[width=.9\textwidth, clip=true]{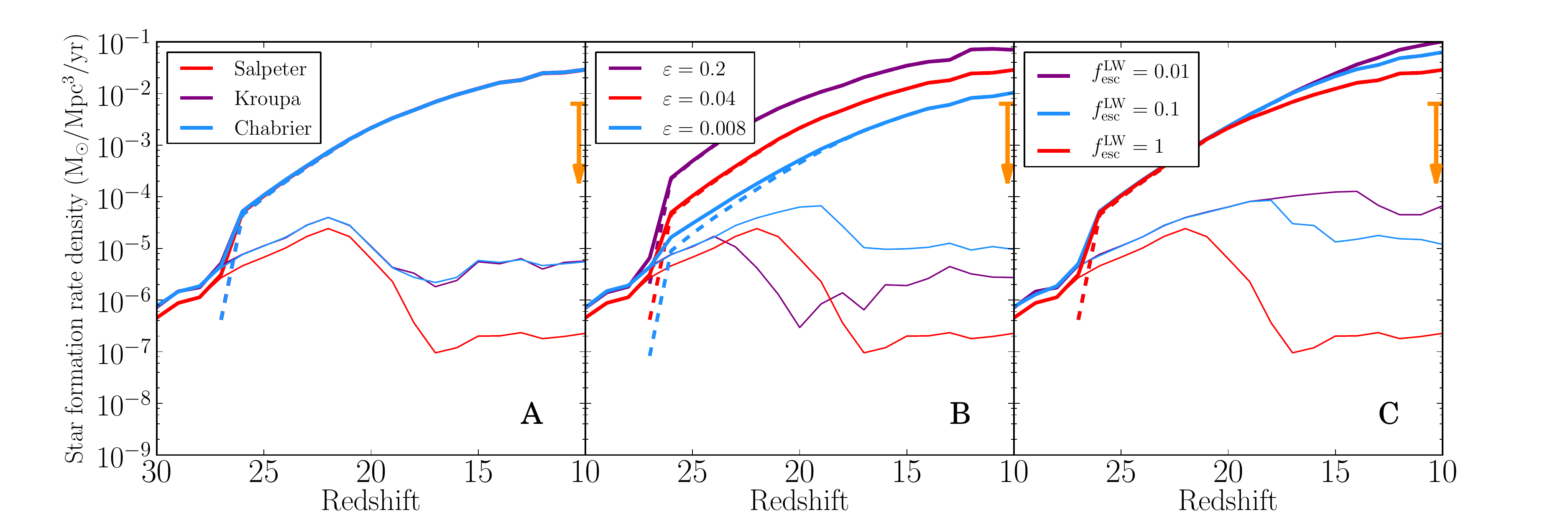} 
   \caption{The star formation rate (SFR) density in M$_\odot$yr$^{-1}$Mpc$^{-3}$ as a function of redshift for variations in IMF, star formation efficiency, and Lyman-Werner (LW) photon escape fraction.  Panel A shows the effect of varying the IMF, Panel B shows the effect of varying the star formation efficiency, and Panel C shows the effect of varying the LW photon escape fraction.  In all panels, the Population III SFR density is plotted in thin, solid lines, the chemically enriched SFR density is plotted in dashed lines, and the total SFR density is plotted in thick, solid lines.  An extrapolated observational upper limit from \citet{2011ApJ...737...90B} is shown in orange.  In all cases, chemically enriched star formation rapidly dominates Population III star formation by several orders of magnitude, but Population III star formation continues at low levels for the entirety of the simulation.}
   \label{fig:5Mpc_sfr_grid}
\end{figure*}

Figure \ref{fig:5Mpc_sfr_grid} shows variations from the canonical model, with different IMFs in panel A, different star formation efficiencies in panel B, and different values of $f^{\rm{LW}}_{\rm{esc}}$ in panel C.  The results in Figure \ref{fig:5Mpc_sfr_grid} are very similar to those shown for the $7.0~h^{-1}$Mpc simulation in Figure \ref{fig:sfr_grid}.  In Panel A, the total and component SFR densities for the three IMFs are nearly indistinguishable.  Panel B shows the SFR density using a Salpeter IMF and varying the star formation efficiency.  Increasing the star formation efficiency increases the difference between the Population III and chemically enriched SFR densities, driving Population III star formation down and chemically enriched star formation up.  Panel C shows the most pronounced difference: the dependence on $f^{\rm{LW}}_{\rm{esc}}$ of the chemically enriched SFR density, where higher values of $f^{\rm{LW}}_{\rm{esc}}$ lead to lower chemically enriched SFR densities.  Increasing $f^{\rm{LW}}_{\rm{esc}}$ increases $J_{21}$, raising minimum halo mass for Population III star formation.  When $J_{21}$ is sufficiently high it will prevent Population III star formation in halos that would have been sufficiently massive to host star formation had the photodissociating radiation been less, as can be seen in the downturns in the Population III SFR densities between $z=22$ and $z=14$.  This downturn occurs at higher redshift with greater $f^{\rm{LW}}_{\rm{esc}}$.  When this happens, fewer halos form Population III stars and become enriched, in turn providing fewer sites for chemically enriched star formation.  This can be seen in panel C of Figure \ref{fig:sfr_grid} in the divergences in the Population III SFR densities at $z=22$ and $z=18$.  In each case, following the drop in the Population III SFR density the chemically enriched SFR density decreases in comparison the models with lower values of $f^{\rm{LW}}_{\rm{esc}}$ as there are now fewer halos capable of hosting chemically enriched star formation.

\begin{figure*}[htbp] 
   \centering
   \includegraphics[width=.9\textwidth, clip=true]{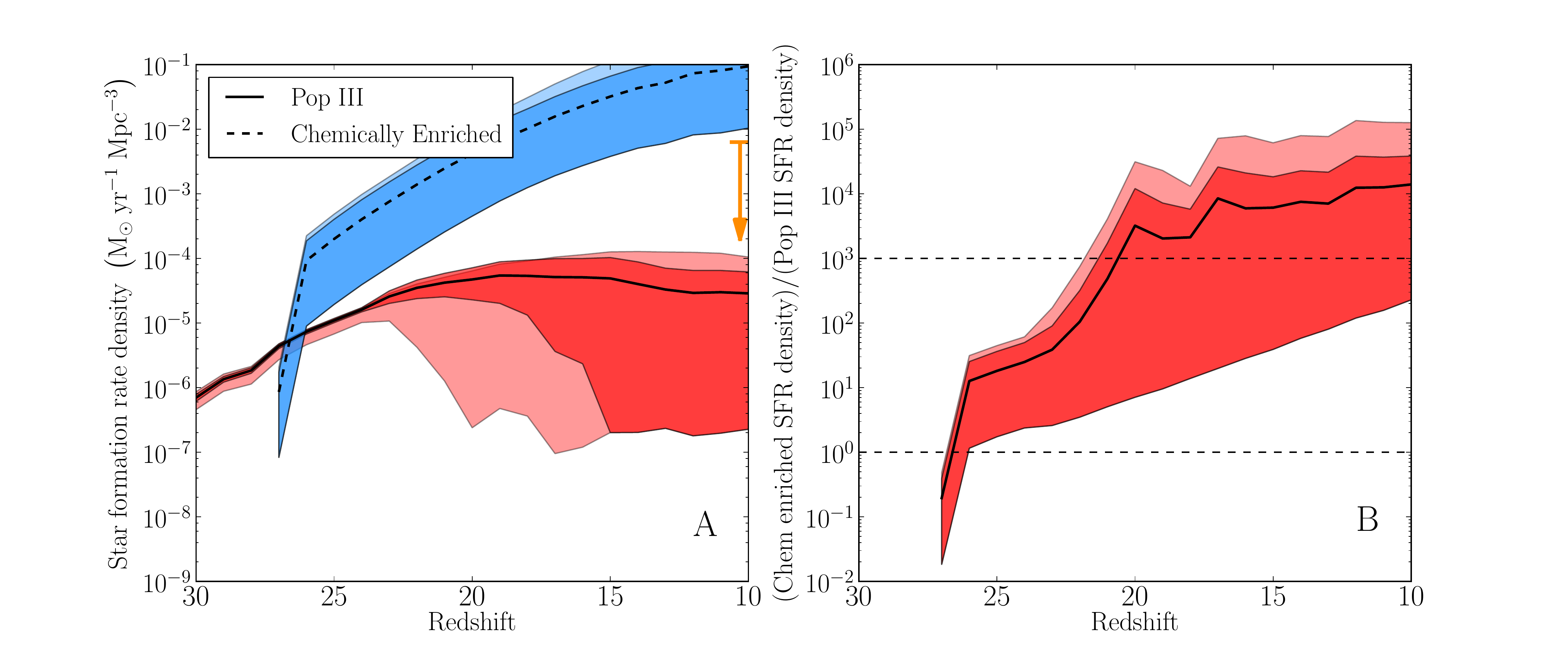} 
   \caption{Same as Figure \ref{fig:sfr_range}, though applied to a $3.5~h^{-1}$Mpc box.  Panel A shows the variation in SFR density in M$_\odot$yr$^{-1}$Mpc$^{-3}$ as a function of redshift for all combinations of parameters in our model.  The mean Population III SFR density is plotted as a black solid line, and the average chemically enriched SFR density is plotted as a black dashed line.  The maximum range spanned by the Population III and chemically enriched SFR densities are shown by the light red and blue shaded regions, respectively.  The dark shaded regions show the 68 percent confidence intervals around the mean.  An extrapolated observational upper limit from \citet{2011ApJ...737...90B} is shown in orange.  Panel B shows the ratio of the chemically enriched SFR density to the Population III SFR density as a function of redshift for all parameter combinations, with the black line showing the average value and the shaded regions having the same meaning as in Panel A.  To ease viewing, dashed lines are shown at 1 and 1000.}
   \label{fig:5Mpc_sfr_range}
\end{figure*}

Comparing the SFR densities found in the $3.5~h^{-1}$Mpc box to those found in the $7.0~h^{-1}$Mpc box shows good agreement, with the Population III SFR density using our fiducial parameters at $z=10$ differing by only $37\%$, and the chemically enriched SFR density differing by $9.4\%$.  The mean Population III SFR density at $z=10$ in the $3.5~h^{-1}$Mpc box is $2.9\times10^{-5}~$M$_\odot$yr$^{-1}$Mpc$^{-3}$, and the mean chemically enriched SFR density is $9.4\times10^{-2}~$M$_\odot$yr$^{-1}$Mpc$^{-3}$.  The most extreme sets of model parameters produce a minimum Population III SFR density of $2.7\times10^{-6}~$M$_\odot$yr$^{-1}$Mpc$^{-3}$, and a maximum of $1.1\times10^{-4}~$M$_\odot$yr$^{-1}$Mpc$^{-3}$.  The chemically enriched SFR density spans from a minimum of $1.0\times10^{-2}~$M$_\odot$yr$^{-1}$Mpc$^{-3}$ to a maximum of $0.36~$M$_\odot$yr$^{-1}$Mpc$^{-3}$.  The evolution of the Population III and chemically enriched SFR densities is also similar to that in the larger simulation volume, with the chemically enriched SFR density increasing rapidly once star formation in chemically enriched halos begins.  The chemically enriched SFR density surpasses the Population III SFR density by $z=27$.  Figure \ref{fig:5Mpc_sfr_range} is identical in format to Figure \ref{fig:sfr_range}, but shows the SFR densities for the $3.5~h^{-1}$Mpc box.  The thick black and dashed lines are the average Population III and chemically enriched SFR densities, respectively, while the dark shaded regions shows the $68\%$ confidence intervals the and the light shaded show the extent of the maximum and minimum values for all models.  The similarities in qualitative behavior and SFR densities between the $3.5~h^{-1}$Mpc box and $7.0~h^{-1}$Mpc box suggest that the $7.0~h^{-1}$Mpc box is capturing the star formation behavior well and that the classification as halos of objects that are significantly below the minimum halo mass for Population III star formation does little to change the overall character of star formation in this model.  The merging of chemically pristine low mass halos to form a halo sufficiently massive to host Population III star formation is indistinguishable from the formation of the same halo from particles that were not previously considered to be members of a halo.  The merging of a low mass, chemically pristine halo with a chemically enriched halo that is already forming stars is indistinguishable from the chemically enriched halo accreting that same amount of unbound pristine material.  The classification of low mass groups of pristine material as a halo has a negligible impact on the nature of star formation once the mass threshold for Population III star formation has exceeded the halo mass resolution, which happens relatively rapidly.

\subsection{Halo Environment}
\label{sec:halo_environment}
\begin{figure*}[htbp] 
   \centering
   \includegraphics[width=1\textwidth, clip=true]{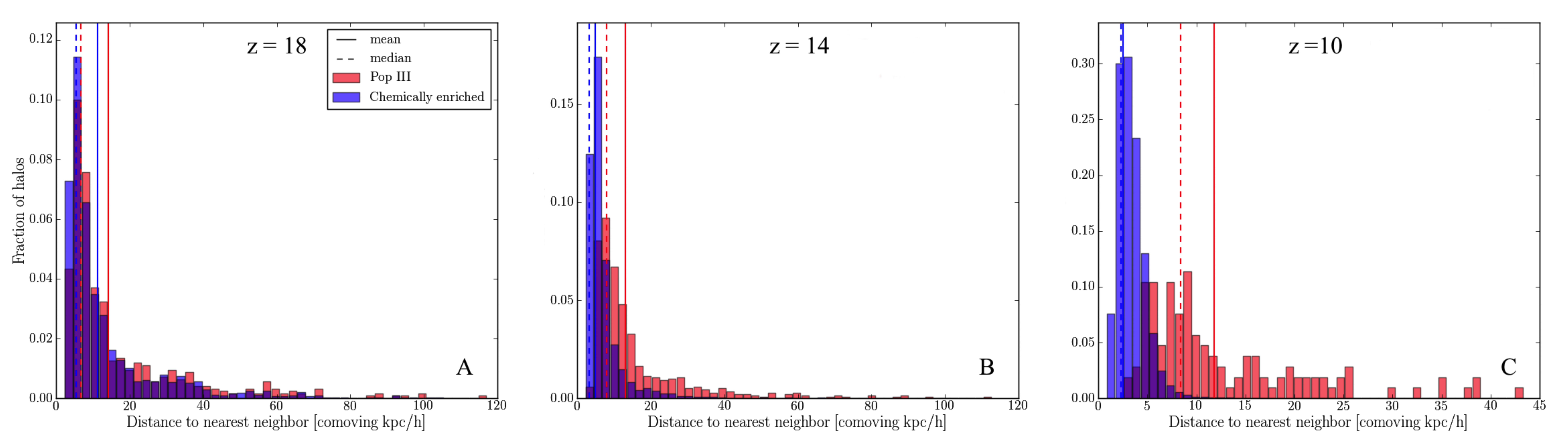} 
   \caption{The distance to the nearest neighboring halo, originating from Population III (red) and chemically enriched (blue) halos.  Panel A shows the distribution of distances at $z=18$, Panel B shows the distribution at $z=14$, and Panel C shows the distribution at $z=10$.  At early times the environments are almost indistinguishable, but as time passes the chemically enriched halos become more clustered and Population III forming halos become increasingly spread out.  The histograms are normalized to allow for the comparison of the much more numerous set of chemically enriched halos to the set of Population III halos.}
   \label{fig:polluted_neighbors}
\end{figure*}

One of the goals of this study is to determine if there was a difference in the local environment of chemically enriched halos and those halos that are chemically pristine and sufficiently massive to form a Population III star.  The distance to the nearest halo is a proxy for the local overdensity and thus environment.  While there are many ways to quantify local overdensity, such as the overall mass overdensity integrated out to some comoving radius, or to some multiple of the virial radius, the distance to the nearest neighboring halo is the most clearly defined, and thus the clearest metric.  Figure \ref{fig:polluted_neighbors} shows the distance to the nearest neighboring halo for our fiducial model for Population III and chemically enriched halos at $z=18$, $14$, and $10$.  At earlier times the distribution of distances to the nearest neighbors is nearly identical for the Population III and chemically enriched halos, suggesting that these halos form in environments that are comparably dense.  As time progresses the chemically enriched halos become much more concentrated (i.e., chemically enriched star formation occurs in progressively denser environments), with nearly no halos more than $10$ comoving $h^{-1}$kpc from their nearest neighbor.  While the chemically enriched distribution becomes more peaked, the Population III distribution becomes broader, suggesting that Population III stars form in halos that are in progressively less dense environments.  This reflects the intuitive expectation that a reservoir of pristine gas large enough to form a Population III star is more likely to assemble and exist in a region that is further away from chemically enriched halos, as even a single merger with a chemically enriched halo will render it incapable of forming a Population III star.  That both distributions have lower maximum distances to the nearest neighboring halo reflects the general growth of cosmic structure and the tendency for objects to be more clustered at lower redshifts.

\begin{figure*}[htbp] 
   \centering
   \includegraphics[width=1\textwidth, clip=true]{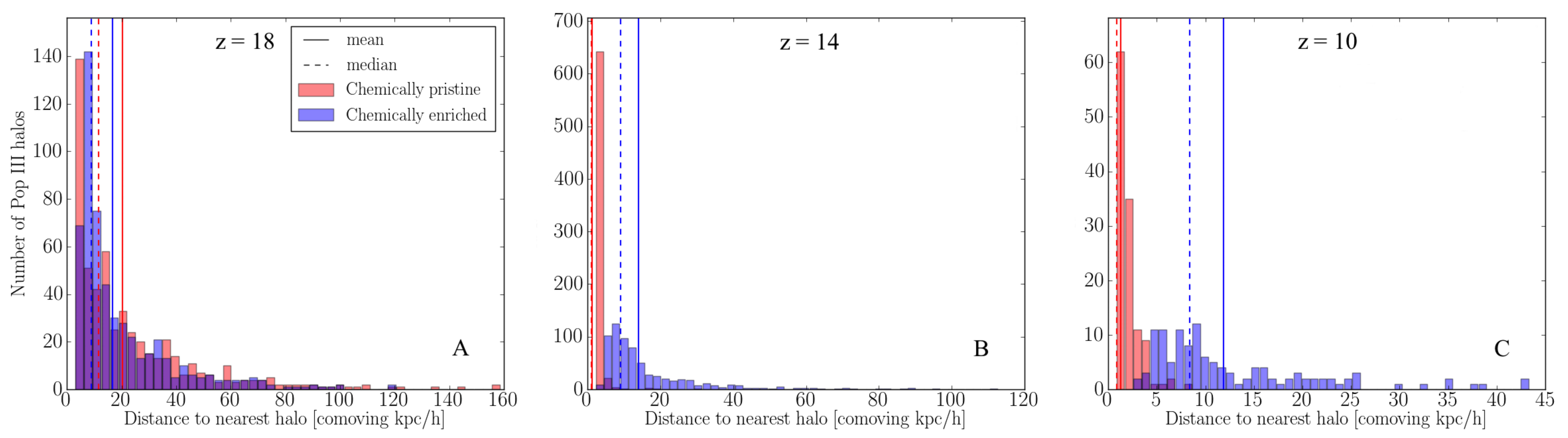} 
   \caption{The distance from halos forming Population III stars to the nearest chemically pristine halo of any mass (red) and to the nearest chemically enriched halo of any mass (blue).  The chemically pristine halos do not need to be massive enough to form a Population III star.  Panel A shows the distribution of distances at $z=18$, Panel B shows the distribution at $z=14$, and Panel C shows the distribution at $z=10$.  Halos hosting Population III star formation are much closer to other chemically pristine halos than to chemically enriched halos.}
   \label{fig:neighbors}
\end{figure*}

The chemical enrichment of the environment of Population III star forming halos can be quantified by comparing the distances from a Population III star forming halo to the nearest chemically pristine halo of any mass (including those below the threshold mass for Population III star formation) and the distance from a Population III star forming halo to the nearest chemically enriched halo.  Figure \ref{fig:neighbors} shows that halos that form Population III stars are generally much closer to other pristine halos than they are to chemically enriched halos.  This agrees with the indications from Figure \ref{fig:polluted_neighbors} that Population III halos will tend to form in low-density regions that are not yet polluted due to the lack of previous star formation in the area, as opposed to high-density regions that typically host chemically enriched star formation.  The assembly of a halo massive enough to form a Population III star in such an environment is typically made possible by the presence and subsequent rapid merger of several smaller, chemically pristine halos that are separately below the threshold for primordial star formation.

\begin{figure*}[htbp] 
   \centering
   \includegraphics[width=1\textwidth, clip=true]{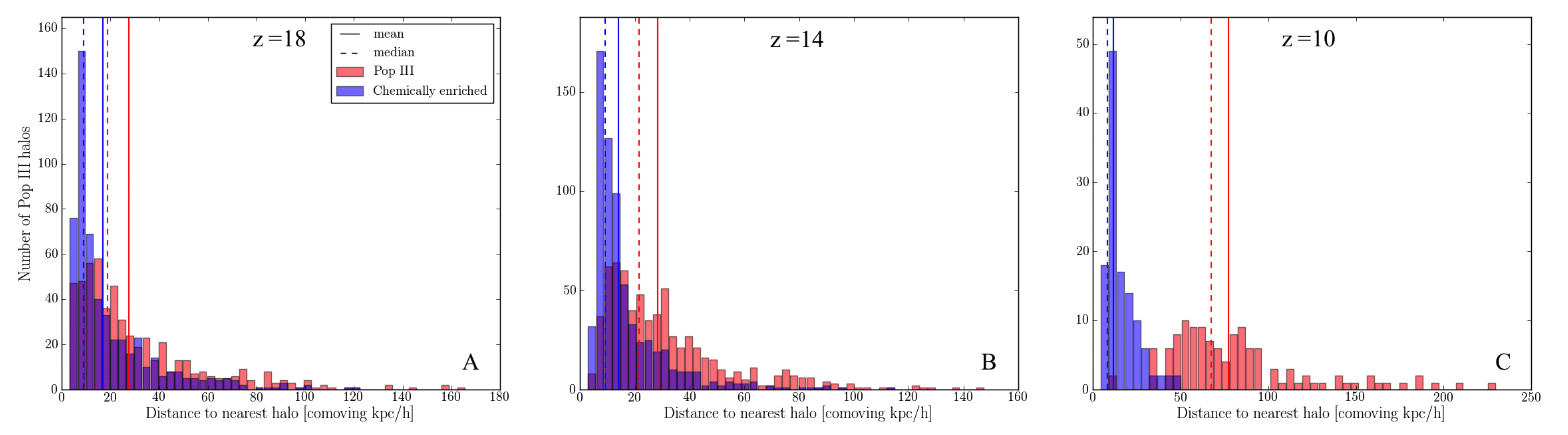} 
   \caption{The distance from Population III star forming halos to the nearest other Population III star forming halo (red) and nearest chemically enriched halo of any mass (blue).  Panel A shows the distribution of distances at $z=18$, Panel B shows the distribution at $z=14$, and Panel C shows the distribution at $z=10$.  The nearest star forming halos are almost entirely chemically enriched, and Population III star forming halos tend to form in isolation from one another.  Taken with Figure \ref{fig:neighbors}, this plot indicates that Population III star forming halos are generally surrounded by chemically pristine halos that are not sufficiently massive to form a Population III star.}
   \label{fig:stellar_neighbors}
\end{figure*}

To further understand the environment in which Population III stars form, Figure \ref{fig:stellar_neighbors} shows the distance from halos massive enough to form a Population III to the nearest other halo that is also massive enough to host Population III star formation, and separately to the nearest chemically enriched halo of any mass.  This shows that Population III stars tend to form in halos that are isolated from one another, as they are almost always more likely to be nearer to a chemically enriched halo than they are to another Population III star forming halo.  Figures \ref{fig:neighbors} and \ref{fig:stellar_neighbors}, taken together, indicate that a Population III star forming halo is most likely to form in a region of many small, chemically pristine halos that are below the mass threshold for Population III star formation.  These low mass chemically pristine halos merge until a halo sufficiently massive to host a Population III star has assembled.  The resulting halo that has surpassed the mass threshold for Population III star formation will be surrounded by the chemically pristine halos that have not yet merged with it and which are incapable of star formation, resulting in Population III stars forming in environments where the nearest neighbors are chemically pristine, smaller halos.  After the Population III star has formed, subsequent mergers of the low mass, chemically pristine halos with this post-Population III halo will fuel chemically enriched star formation.  As the Population III star forming halo was the most massive local object, this effectively shuts off Population III star formation in this region.

\begin{figure*}[htbp] 
   \centering
   \includegraphics[width=1\textwidth, clip=true]{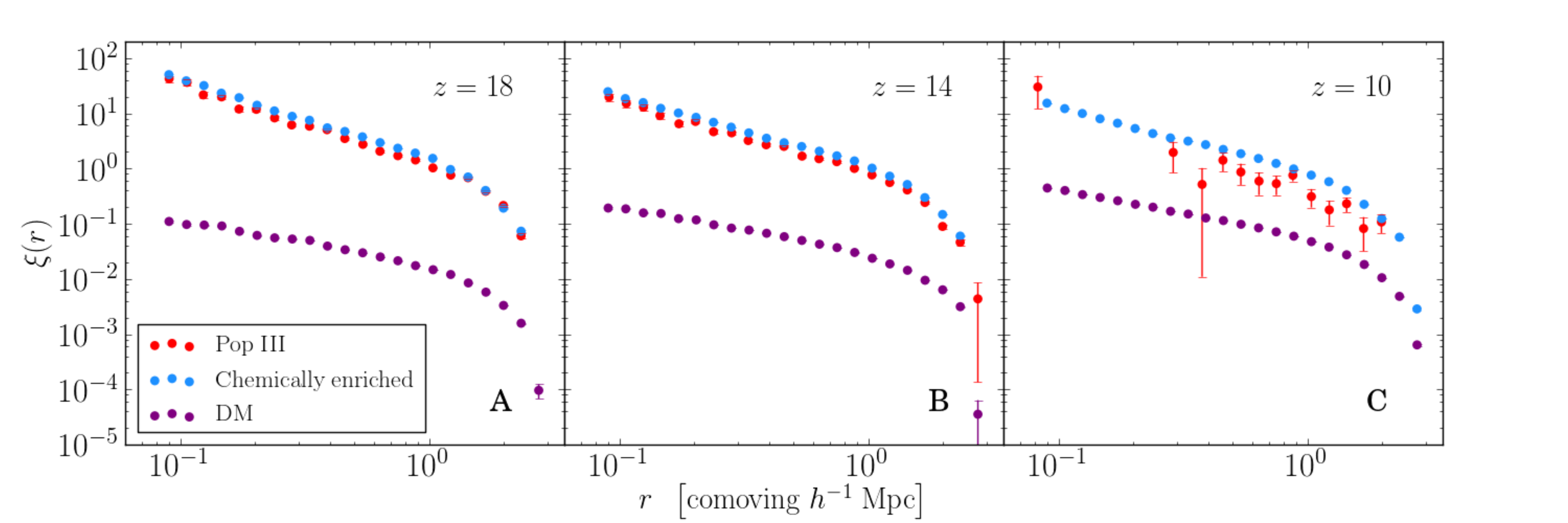} 
   \caption{The halo correlation functions for Population III halos (red) and chemically enriched halos (blue), and the unbiased dark matter density field (violet).  Panel A, B, and C show the correlation functions at $z=18$, $14$, and $10$, respectively.  Error bars are plotted for all points, but are generally not visible.  At all times chemically enriched halos are more clustered than Population III star forming halos.}
   \label{fig:correlation_function}
\end{figure*}

We next investigate the spatial distribution of the Population III star forming halos and the chemically enriched halos using two-point statistics. The two-point correlation function $\xi( r)$ quantifies the excess probability (w.r.t. random) to find two halos in volume elements $dV_1$ and $dV_2$ separated by distance $r$,
\begin{equation}
dP_{12}(r ) = \bar{n}^2\left[ 1+\xi( r) \right] dV_{1}dV_{2},
\end{equation}
where $\bar{n}$ is the mean number density of halos. We compute the $\xi(r)$ using the estimator introduced by \citet{1993ApJ...417...19H}.  We complement the (biased) halo correlation functions with the correlation functions of the unbiased dark matter density field, which we estimate from 100,000 randomly drawn dark matter particles from the simulations. The results of our clustering analysis are given for three redshifts in Figure \ref{fig:correlation_function}, where we show the correlation functions of Population III star forming halos (red), chemically enriched halos (blue), and the dark matter density field (violet). The decrease in the correlation functions at scales larger than approximately half the box size ($\sim3.5~h^{-1}$Mpc) is due to the finite size of the simulation volume.  At all times, the chemically enriched halos are slightly more clustered than the Population III star forming halos. We estimate the bias of Population III star forming halos to evolve from an extreme value of $\sim 10$ to $\sim 5$ to $\sim 3$ from $z=18$ to $14$ to $10$. This evolution is complemented by the growth of a pronounced exclusion region in the two-point correlation function of Population III halos that becomes clearly visible by $z=10$; no Population III star forming pairs are found at separations below $8.2$ comoving $h^{-1}$kpc, consistent with the nearest neighbor analysis discussed above and shown in Figures \ref{fig:polluted_neighbors}-\ref{fig:stellar_neighbors}.

\begin{figure}[htbp] 
   \centering
   \includegraphics[width=.45\textwidth, clip=true]{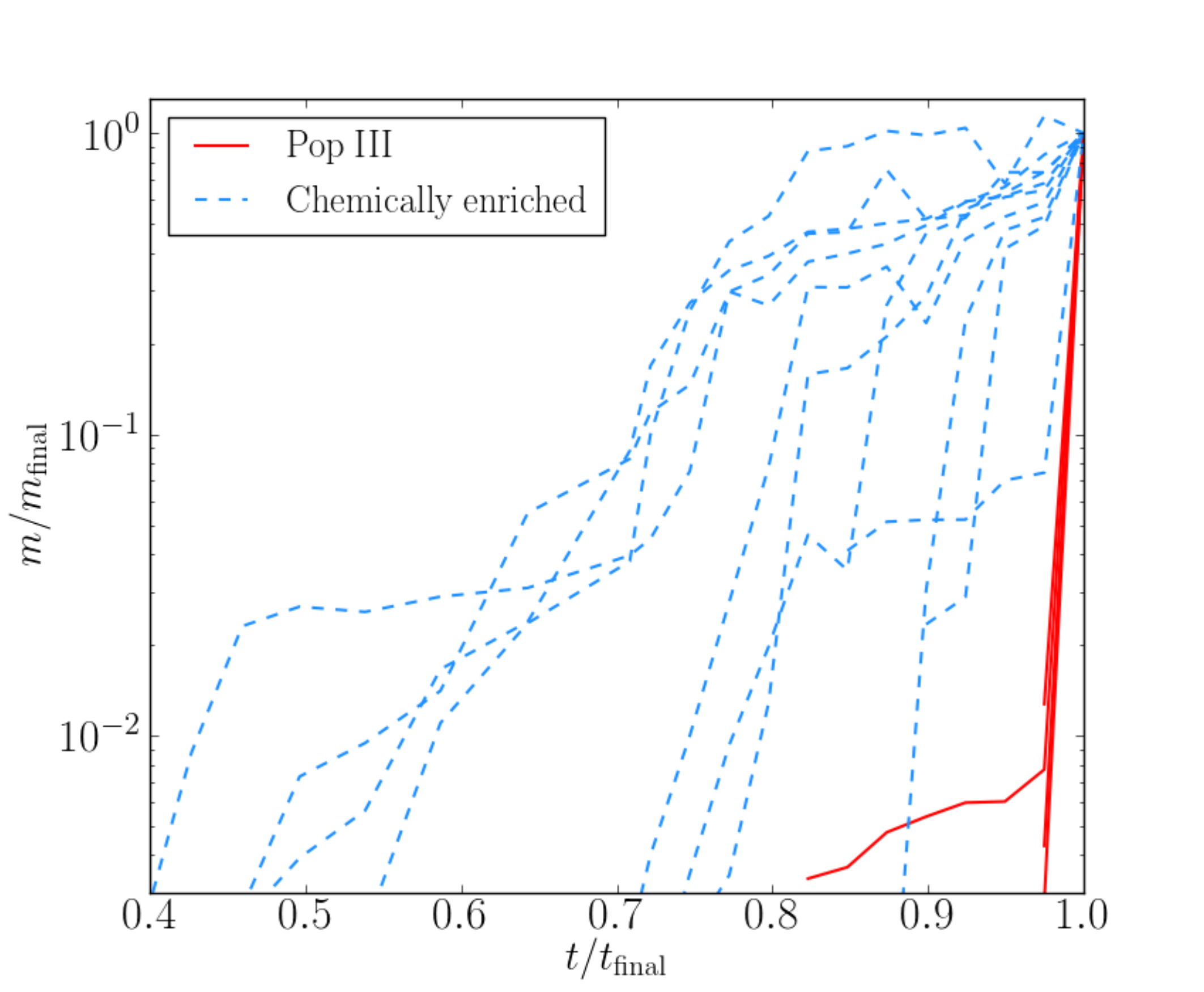} 
   \caption{An example of the rate of growth of halos to their final mass at $z=10.73$.  Each halo is normalized to its final mass, and each line represents an individual halo.  Time is shown on the horizontal axis, with $t_\mathrm{final}$ being defined as the amount of time since the first star in the simulation formed.  Red, dashed lines show the 10 most massive chemically enriched halos and blue, solid lines show the 10 most massive chemically pristine halos.  All 10 chemically pristine halos are plotted, though their very similar growth at late times makes them overlap in this figure.  Chemically enriched halos experience faster growth at early times, while the chemically pristine halos that host Population III stars grow slowly at early times, remaining below the mass required for star formation.  Growth in chemically pristine halos occurs rapidly at late times, immediately prior to star formation.}
   \label{fig:halo_growth}
\end{figure}

The nature of Population III star forming halo assembly at late times can be better understood by examining the halo merger history that precedes primordial star formation in a typical halo.  Figure \ref{fig:halo_growth} shows a comparison of several halos that host Population III stars to several halos that form chemically enriched stars.  The halo lineage is followed backwards through time, and follows the most massive progenitor.  Figure \ref{fig:halo_growth} works backwards in time from $z=10.73$, though similar behavior is seen in across late times in our model.  The mass of each halo in the lineage is normalized to the final mass of the halo, and the time is normalized to the time elapsed in the simulation since the first star formed.  There is a clear difference in the growth of the two populations of halos.  Halos that host chemically enriched stars grow more rapidly at early times, with slower growth later.  This pattern of halo growth in overdense regions, in which rapid early growth is followed by slower growth at later times, is in agreement with the findings of \citet{2009MNRAS.398.1858M}, and supports our finding that chemically enriched star formation occurs in halos that are in more overdense regions.  Conversely, chemically pristine halos grow slowly at early times, and experience rapid mergers at late times that push their mass above the threshold required for Population III star formation.  The rapid, late assembly of Population III halos at later times in our models is the result of multiple mergers of small, chemically pristine halos occurring in rapid succession.  This rapid growth, taken in the context of Figures \ref{fig:neighbors} and \ref{fig:stellar_neighbors}, shows a coherent picture of a region of many small, chemically pristine halos, all of which are individually too small to form a Population III star, undergoing a rapid series of mergers until the Population III star formation mass threshold is surpassed.  When the Population III star forms, its host halo is still surrounded by chemically pristine halos, but subsequent mergers with this now chemically enriched halo fuel chemically enriched star formation rather than Population III star formation.  Halos in underdense regions grow as if they are in a low-$\Omega_{M}$ universe, and are thus retarded compared to the mean \citep{1996MNRAS.282..347M}.  This explains the slow growth and systematically different behavior.  The slow growth of halos in these underdense regions enables the formation of chemically pristine halos large enough to host Population III star formation at late times according to our models.

\section{Discussion}
\label{sec:discussion}
\subsection{Comparison to Observation}
To verify that this model accurately creates a set of halos capable of forming Population III stars, several methods of validation were undertaken.  One method was to compare the total star formation rate (SFR) density to observational constraints from \citet{2011ApJ...737...90B}.  The stellar mass density derived SFR density they report for $z=10.3$ ranges from approximately $2.5\times10^{-3}$ to $6.3\times10^{-3}~$M$_\odot$yr$^{-1}$Mpc$^{-3}$, with a luminosity density derived SFR density placing an upper limit at $2.5\times10^{-4}~$M$_\odot$yr$^{-1}$Mpc$^{-3}$.  This range brackets the SFR density found by our model with a star formation efficiency of $0.008$, and is in reasonable agreement with the models with higher star formation efficiencies.  It should be noted that the SFR density reported in \citet{2011ApJ...737...90B} at $z=10.3$ is an extrapolation from data extending to $z\sim8$, and in the case of the luminosity function derived SFR density is reported as an upper limit.

\subsection{Comparison to Other Work}
The work of \citet{2009ApJ...694..879T} motivated the Population III star formation method used in this model, and comparison with their results shows a striking agreement.  While \citet{2009ApJ...694..879T} use an analytic dark matter halo formation rate derived from the \citet{1999MNRAS.308..119S} mass function, as opposed to the cosmological simulations used here, there is excellent agreement between the two works.  The minimum halo mass for Population III star formation at $z=10$ is nearly identical: $4.8\times10^7~$M$_\odot$ with our fiducial model compared to approximately $6.4\times10^7~$M$_\odot$ by \citet{2009ApJ...694..879T}.  The SFR density at $z=10$ also demonstrates good agreement, with \citet{2009ApJ...694..879T} finding approximately $4\times10^{-3}~$M$_\odot$yr$^{-1}$Mpc$^{-3}$, compared to $1.3\times10^{-2}~$M$_\odot$yr$^{-1}$Mpc$^{-3}$ with our fiducial model.  This level of agreement is encouraging considering that our model utilizes a cosmological simulation to determine halo populations rather than the analytic \citet{1999MNRAS.308..119S} mass function, our Population III star formation model was modified, and that our chemically enriched star formation model was entirely different than the fixed rate of \citet{2009ApJ...694..879T}.  Given that chemically enriched star formation dominates the overall star formation at $z=10$ in both models, this is particularly encouraging.  

This work compares well with work by \citet{2002ApJ...575...49R}, who find a SFR density of approximately $2\times10^{-2}~$M$_\odot$yr$^{-1}$Mpc$^{-3}$ at $z=10$ despite vast differences in approach.  The simulations of \citet{2002ApJ...575...49R} self-consistently solve the radiative transfer equation, utilize H$_2$ chemistry, heating, and cooling networks, as well as a Schmidt law star formation prescription.  \citet{2002ApJ...575...49R} use much smaller volumes of $0.5$, $1$, and $2$ comoving $h^{-1}$Mpc on a side, giving considerably less statistical power to their results.  The advantage to their work is the addition of much more elaborate multiphysics processes governing star formation and radiative feedback.  

The work of \citet{2012ApJ...745...50W} is similarly much more robust in its multiphysics capabilities, though is again hampered by a small simulation volume of $1~h^{-1}$Mpc$^3$.  The simulations of \citet{2012ApJ...745...50W} use 12 levels of adaptive mesh refinement, a nine-species non-equalibrium chemistry network, prescriptions for both Population III and chemically enriched star formation, as well as kinetic and radiative stellar feedback.  Regardless, our model finds very similar values of the SFR density, both qualitatively and quantitatively, for both Population III and chemically enriched star formation (see their Figure 3).  Our models utilizing $f_{\rm{esc}}^{\rm{LW}}=0.01$ and $0.1$ are in particularly good agreement with their findings.  Furthermore, in both our results and those of others, the chemically enriched SFR density rises to dominate the Population III SFR density by several orders of magnitude by $z=10$.

\subsection{Implications for Population III Modeling}
Treating the photodissociating radiation self-consistently is crucially important to the selection of halos that are capable of forming a Population III star.  Figure \ref{fig:j21_and_m_threshold} indicates that the destruction of H$_2$ dramatically impacts the minimum halo mass that is capable of hosting Population III star formation.  As Population III star formation persists well beyond the formation of the first star in the simulation, the vast majority of Population III stars will form in the presence of a non-negligible photodissociating background.  These stars, termed Population III.2 \citep{2008ApJ...681..771M,2008AIPC..990D..13O}, comprise the majority of stars that form from chemically pristine gas, but are fundamentally impacted by the radiation produced by other stars.  This suggests that the simulations that use small simulation volumes (e.g., \citet{2002Sci...295...93A,2002ApJ...564...23B,2009Sci...325..601T}) are typically neglecting a major aspect of the environment in which Population III stars form.  Even some of those that do include this effect (e.g., \citet{2001ApJ...548..509M,2008ApJ...673...14O,2003ApJ...592..645Y}) suffer from the effects of small simulation volumes and the the lack of metal-enriched stellar populations, and simulations that include both (e.g., \citet{2002ApJ...575...33R,2012ApJ...745...50W}) still have too small of a simulation volume to adequately sample the star formation behavior of the early universe.

\subsection{Limitations and Future Work}
The photodissociating background is treated as being homogenous in the entire simulation volume rather than preferentially impacting the halos nearest to star forming halos.  On the small scale, one expects the photodissociating background to vary as $r^{-2}$, where $r$ is the distance from nearest small number of halos, and should vary substantially.  We could in principle account for this variation on small scales, though the long-lived photodissociating background can be modeled as being the sum of a homogeneous term and an anisotropic term.  The anisotropic term is caused by star formation in the last approximately $L_{\rm{box}}/c(1+z)$ years, where $L_{\rm{box}}$ is the size of the simulation volume and $c$ is the speed of light.  On a larger scale, the photodissociating background is inhomogeneous on the many-Mpc scale based on large-scale modes \citep{2009ApJ...695.1430A}.  This inhomogeneity is neglected as it is neither a large effect in comparison to the homogeneous photodissociating background, and because it cannot be treated in a model of this nature.

The effect of ionizing radiation produced by the stellar populations is neglected.  The recombination rate scales as $\rho^2$, where $\rho$ is the density, which in turn scales as $(1+z)^3$, so the recombination rate will scale as $(1+z)^6$.  At the redshifts of interest, halos would need to be very close to one another for the ionized hydrogen regions to be important.  For example, calculating the radius of the Str\"{o}mgren sphere produced by a 3 M$_\odot$ chemically enriched star using the average ionizing flux from \citet{2003A&A...397..527S} and an average density approximating the hydrogen mass density component of the virial density, calculated as $178\times(\Omega_{\rm{B}}\rho_{\rm{c}}(1+z)^3)/(m_{\rm{H}})$, where $\rho_{\rm{c}}$ is the critical density of the universe, and $m_{\rm{H}}$ is the mass of a hydrogen atom, gives an ionized region around the star extending only $15$ pc at $z=30$, and only $122$ pc at $z=10$.

Any gas that is ejected from a halo is considered to be permanently lost, and is never incorporated into future halos or used in star formation.  As the mass of gas that is lost is determined by the energetics of the supernova in the halo and the escape velocity as determined by the halo mass, the assumption that the material is permanently lost from its halo of origin is likely valid.  \citet{2005ApJ...630..675K} show that the fate of gas in a halo in which a supernova occurs cannot be determined solely through the comparison of the explosion energy and the binding energy of the halo, but is strongly dependent on gas density profile, and that in halos of mass $\sim10^7$ M$_\odot$ and larger the halo will not be evacuated, even when the explosion energy exceeds the binding energy by 2 orders of magnitude.  Chemically enriched gas could conceivably be ejected from one halo and impinge on a nearby pristine halo, rendering that halo incapable of forming a Population III star despite never having hosted star formation in its assembly history.  While chemically enriched material may be ejected from a halo as a result of a supernova explosion, it extends to a radius of only $\sim1~$kpc within $10^5-10^7$ yrs \citep{2003ApJ...596L.135B,2008ApJ...682...49W}, and has a negligible impact on star formation in satellite minihalos \citep{2010ApJ...712..101W}.  It is unlikely that chemically enriched material ejected into the intergalactic medium would pollute surrounding chemically pristine halos, as at $z=10$ the minimum comoving separation between a Population III and chemically enriched halo is $2.2~h^{-1}$kpc. These motivations are at the core of the assumption that star formation, evolution, and death in one halo does not directly impact other halos that are nearby, but only contribute to the global characteristics of the simulation volume.  

This model does not include any effects of reionization, but this is likely not an issue as the simulations to which the model is applied are stopped at $z=10$, well before the epoch of reionization.  The results of this model will be invalid if applied to times beyond the onset of the epoch of reionization.  

Future work allows several primary areas for improvement in this model.  Improved methods for associating dark matter particles with a specific halo based on the gravitational potential, or by utilizing a six-dimensional Friends of Friends algorithm \citep{2006ApJ...649....1D} that includes particle velocity data in addition to the spatial criteria for halo identification would provide a more physically realistic halo catalog and merger tree.  The very rapid growth of Population III halos at late times (see Figure \ref{fig:halo_growth}) has potentially important implications for the formation of Population III stars.  \citet{2007ApJ...654...66O} showed that increasing the halo merger rate, and in turn growth rate, increased the temperature of the halo, leading to the production of more H$_2$, though the results of \citet{2007ApJ...654...66O} are for much larger mass halos, so care should be taken in extending their results to the halos in our simulation.  Increasing the H$_2$ content of a halo could lead to more efficient cooling, and a colder halo core.  The method for determining if a halo is capable of forming a Population III star in this model is independent of the halo assembly history, and uses halo mass as a proxy for the maximum amount of H$_2$ that can reside in the halo.  Allowing the model to increase the mass of H$_2$ in a halo in response to rapid growth could potentially enable halos less massive than the current Population III star forming mass threshold to cool efficiently and form stars.  Treatment and addition of this effect to the model are left for future work.

Many other effects that are not included in this model may be important to the Population III star formation and the determination of the Population III initial mass function (IMF).  For example, molecular hydrogen rates \citep{2011ApJ...726...55T}, magnetic fields \citep{2012ApJ...745..154T}, and subgrid turbulence \citep{2012arXiv1212.1619L} have been shown to be dynamically important in Population III star formation.  Turbulence in primordial clouds enhances fragmentation, even when subsonic, and this behavior is observed in both Population III.1 and Population III.2 star forming halos \citep{2011ApJ...727..110C}.  Radiative feedback from a Population III star can halt accretion, establishing its mass, and H$_2$ photodissociating radiation could reduce the differences mass between Population III.1 and Population III.2 stars by inhibiting cooling via H$_2$ \citep{2012ApJ...760L..37H}.  Investigations using sink particles to model the accreting protostars have indicated that local radiative feedback halts accretion onto the protostars forming in a fragmented primordial disk, and that the end of accretion is what sets the final stellar masses \citep{2010MNRAS.403...45S,2012MNRAS.422..290S}.  The impact of additional energy input from dark matter annihilation on the fragmentation properties of a Population III star forming cloud is investigated by \citet{2012ApJ...761..154S}, who find that fragmentation still occurs despite this additional energy.  It may be the case that the Population III IMF is set primarily by effects on a scale smaller than the halo environment, and that large scale differences in environment are a sub-dominant effect, but this investigation must be carried out in order to assess the importance of halo environment on Population III star formation.

The results of this model will be improved with its application to larger simulation volumes, as increasingly rare high mass halos will be more likely to form.  This will allow for a more representative investigation of the halos that host both Population III star formation and the feedback that ensues as chemically enriched star formation begins.  The application of this model to simulations run with full physics capabilities will allow for direct comparison between the results of the two methods.  This will enable the identification of physical processes that are relevant to star formation and chemical evolution that cannot be treated in the statistical manner of this model.  Identification and improvement of these areas will allow for the model to become more robust while retaining as much of the current computational efficiency as possible. 

This is the first paper in a series.  Paper II investigates the nucleosynthetic evolution of high redshift structure in comparison to the local dwarf spheroidal population and the Milky Way and Andromeda stellar halos.  Further papers will include full-physics adaptive mesh refinement simulations of selected pristine halos across a range of redshifts and density environments.  

\section{Summary and Conclusions}
\label{sec:conclusions}
Our model identifies the chemically pristine halos capable of forming a Population III star in an N-body cosmological simulation.  The semi-analytical model includes Population III and chemically enriched star formation, halo metal pollution, and the H$_2$ photodissociating radiation from the stellar population.  This is a substantial improvement over previous work of this type, and is also a useful complement to full-physics simulations because it allows for the investigation of larger cosmological volumes, allowing for improved statistics and the creation of a more representative sample of Population III star forming halos.  Population III and chemically enriched star forming halos have very similar properties and environments at high redshifts, but these properties diverge substantially at later times.  At late times Population III stars form in massive halos in underdense regions that grow rapidly.  Population III star forming halos assemble in isolation from both chemically enriched and other Population III star forming halos.  This finding carries implications for the search for Population III stars with the James Webb Space Telescope and other future observational missions, as Population III stars do not form in or near galaxies, and searches in these environments are unlikely to yield results.  

Accurate modeling of Population III star formation requires effects and conditions that are not accounted for in current simulations.  Work to model Population III star formation in the redshift range of $z\sim15$ to $z\sim10$ should be carried out in larger simulation volumes in order to self-consistently determine the H$_2$ photodissociating radiation produced by other stars.  The density environment that hosts a Population III star forming halo is generally modeled improperly as well, suggesting that the Population III star formation in the literature to date may reflect only a portion of the character of that found in nature.

This model has succeeded in producing a catalog of more than 40,000 halos in a single cosmological simulation that are capable of forming Population III stars.  These halos range in mass from $2.3\times10^5~$M$_\odot$ to $1.2\times10^{10}~$M$_\odot$ and in redshift  from $z=30$ to $z=10$.  Simulations of these halos will enable a vastly more representative study of the characteristics of Population III stars.

\section{Acknowledgments}
This work used the Extreme Science and Engineering Discovery Environment (XSEDE), which is supported by National Science Foundation grant number OCI-1053575.  All simulations were funded by XSEDE award TG-AST090040.  In particular, we would like to thanks the RDAV staff at NICS for their support and MSU's Institute for Cyber-Enabled Research for access to computing resources.  This work was funded by the NASA ATFP program (NNX09AD80G and NNX12AC98G), the NSF AST program (AST-0908819), the LANL Institute for Geophysics and Planetary Physics, and MSU's Institute for Cyber-Enabled Research.  M.J.T was supported in this work by NSF CI TraCS fellowship award OCI-1048505.  O.H. acknowledges support from the Swiss National Science Foundation (SNSF) through the Ambizione fellowship.  B.W.O. would like to thank Tom Abel and Michael Norman for useful discussions.  We thank Michele Trenti for helpful discussion of Population III star formation modeling, and Stephen Skory for his technical assistance in the creation of the halo merger trees.

\bibliographystyle{apj}
\bibliography{apj-jour,PopIII}

\begin{thebibliography}{73}
\expandafter\ifx\csname natexlab\endcsname\relax\def\natexlab#1{#1}\fi

\bibitem[{{Abel} {et~al.}(2002){Abel}, {Bryan}, \&
  {Norman}}]{2002Sci...295...93A}
{Abel}, T., {Bryan}, G.~L., \& {Norman}, M.~L. 2002, Science, 295, 93

\bibitem[{{Ahn} {et~al.}(2009){Ahn}, {Shapiro}, {Iliev}, {Mellema}, \&
  {Pen}}]{2009ApJ...695.1430A}
{Ahn}, K., {Shapiro}, P.~R., {Iliev}, I.~T., {Mellema}, G., \& {Pen}, U.-L.
  2009, \apj, 695, 1430

\bibitem[{{Bertschinger}(2001)}]{2001ApJS..137....1B}
{Bertschinger}, E. 2001, \apjs, 137, 1

\bibitem[{{Bigiel} {et~al.}(2011){Bigiel}, {Leroy}, {Walter}, {Brinks}, {de
  Blok}, {Kramer}, {Rix}, {Schruba}, {Schuster}, {Usero}, \&
  {Wiesemeyer}}]{2011ApJ...730L..13B}
{Bigiel}, F., {et~al.} 2011, \apjl, 730, L13

\bibitem[{{Bouwens} {et~al.}(2011){Bouwens}, {Illingworth}, {Oesch},
  {Labb{\'e}}, {Trenti}, {van Dokkum}, {Franx}, {Stiavelli}, {Carollo},
  {Magee}, \& {Gonzalez}}]{2011ApJ...737...90B}
{Bouwens}, R.~J., {et~al.} 2011, \apj, 737, 90

\bibitem[{{Bromm} {et~al.}(2002){Bromm}, {Coppi}, \&
  {Larson}}]{2002ApJ...564...23B}
{Bromm}, V., {Coppi}, P.~S., \& {Larson}, R.~B. 2002, \apj, 564, 23

\bibitem[{{Bromm} {et~al.}(2001){Bromm}, {Ferrara}, {Coppi}, \&
  {Larson}}]{2001MNRAS.328..969B}
{Bromm}, V., {Ferrara}, A., {Coppi}, P.~S., \& {Larson}, R.~B. 2001, \mnras,
  328, 969

\bibitem[{{Bromm} \& {Loeb}(2003)}]{2003Natur.425..812B}
{Bromm}, V., \& {Loeb}, A. 2003, \nat, 425, 812

\bibitem[{{Bromm} {et~al.}(2003){Bromm}, {Yoshida}, \&
  {Hernquist}}]{2003ApJ...596L.135B}
{Bromm}, V., {Yoshida}, N., \& {Hernquist}, L. 2003, \apjl, 596, L135

\bibitem[{{Bryan} \& {Norman}(1997)}]{1997ASPC..123..363B}
{Bryan}, G.~L., \& {Norman}, M.~L. 1997, in Astronomical Society of the Pacific
  Conference Series, Vol. 123, Computational Astrophysics; 12th Kingston
  Meeting on Theoretical Astrophysics, ed. D.~A. {Clarke} \& M.~J. {West}, 363

\bibitem[{{Bryan} \& {Norman}(2000)}]{2000IMA...117..165B}
{Bryan}, G.~L., \& {Norman}, M.~L. 2000, Institute for Mathematics and Its
  Applications, 117, 165

\bibitem[{{Chabrier}(2003)}]{2003PASP..115..763C}
{Chabrier}, G. 2003, \pasp, 115, 763

\bibitem[{{Clark} {et~al.}(2011{\natexlab{a}}){Clark}, {Glover}, {Klessen}, \&
  {Bromm}}]{2011ApJ...727..110C}
{Clark}, P.~C., {Glover}, S.~C.~O., {Klessen}, R.~S., \& {Bromm}, V.
  2011{\natexlab{a}}, \apj, 727, 110

\bibitem[{{Clark} {et~al.}(2011{\natexlab{b}}){Clark}, {Glover}, {Smith},
  {Greif}, {Klessen}, \& {Bromm}}]{2011Sci...331.1040C}
{Clark}, P.~C., {Glover}, S.~C.~O., {Smith}, R.~J., {Greif}, T.~H., {Klessen},
  R.~S., \& {Bromm}, V. 2011{\natexlab{b}}, Science, 331, 1040

\bibitem[{{Diemand} {et~al.}(2006){Diemand}, {Kuhlen}, \&
  {Madau}}]{2006ApJ...649....1D}
{Diemand}, J., {Kuhlen}, M., \& {Madau}, P. 2006, \apj, 649, 1

\bibitem[{{Efstathiou} {et~al.}(1985){Efstathiou}, {Davis}, {White}, \&
  {Frenk}}]{1985ApJS...57..241E}
{Efstathiou}, G., {Davis}, M., {White}, S.~D.~M., \& {Frenk}, C.~S. 1985,
  \apjs, 57, 241

\bibitem[{{Galli} \& {Palla}(1998)}]{1998A&A...335..403G}
{Galli}, D., \& {Palla}, F. 1998, \aap, 335, 403

\bibitem[{{Greif} {et~al.}(2012){Greif}, {Bromm}, {Clark}, {Glover}, {Smith},
  {Klessen}, {Yoshida}, \& {Springel}}]{2012MNRAS.424..399G}
{Greif}, T.~H., {Bromm}, V., {Clark}, P.~C., {Glover}, S.~C.~O., {Smith},
  R.~J., {Klessen}, R.~S., {Yoshida}, N., \& {Springel}, V. 2012, \mnras, 424,
  399

\bibitem[{{Hahn} \& {Abel}(2011)}]{2011MNRAS.415.2101H}
{Hahn}, O., \& {Abel}, T. 2011, \mnras, 415, 2101

\bibitem[{{Hamilton}(1993)}]{1993ApJ...417...19H}
{Hamilton}, A.~J.~S. 1993, \apj, 417, 19

\bibitem[{{Heger} \& {Woosley}(2002)}]{2002ApJ...567..532H}
{Heger}, A., \& {Woosley}, S.~E. 2002, \apj, 567, 532

\bibitem[{{Hockney} \& {Eastwood}(1988)}]{1988csup.book.....H}
{Hockney}, R.~W., \& {Eastwood}, J.~W. 1988, {Computer simulation using
  particles}

\bibitem[{{Hosokawa} {et~al.}(2012){Hosokawa}, {Yoshida}, {Omukai}, \&
  {Yorke}}]{2012ApJ...760L..37H}
{Hosokawa}, T., {Yoshida}, N., {Omukai}, K., \& {Yorke}, H.~W. 2012, \apjl,
  760, L37

\bibitem[{{Iwamoto} {et~al.}(1999){Iwamoto}, {Brachwitz}, {Nomoto},
  {Kishimoto}, {Umeda}, {Hix}, \& {Thielemann}}]{1999ApJS..125..439I}
{Iwamoto}, K., {Brachwitz}, F., {Nomoto}, K., {Kishimoto}, N., {Umeda}, H.,
  {Hix}, W.~R., \& {Thielemann}, F.-K. 1999, \apjs, 125, 439

\bibitem[{{Karakas}(2010)}]{2010MNRAS.403.1413K}
{Karakas}, A.~I. 2010, \mnras, 403, 1413

\bibitem[{{Kitayama} \& {Yoshida}(2005)}]{2005ApJ...630..675K}
{Kitayama}, T., \& {Yoshida}, N. 2005, \apj, 630, 675

\bibitem[{{Kitayama} {et~al.}(2004){Kitayama}, {Yoshida}, {Susa}, \&
  {Umemura}}]{2004ApJ...613..631K}
{Kitayama}, T., {Yoshida}, N., {Susa}, H., \& {Umemura}, M. 2004, \apj, 613,
  631

\bibitem[{{Knebe} {et~al.}(2011){Knebe}, {Knollmann}, {Muldrew}, {Pearce},
  {Aragon-Calvo}, {Ascasibar}, {Behroozi}, {Ceverino}, {Colombi}, {Diemand},
  {Dolag}, {Falck}, {Fasel}, {Gardner}, {Gottl{\"o}ber}, {Hsu}, {Iannuzzi},
  {Klypin}, {Luki{\'c}}, {Maciejewski}, {McBride}, {Neyrinck}, {Planelles},
  {Potter}, {Quilis}, {Rasera}, {Read}, {Ricker}, {Roy}, {Springel}, {Stadel},
  {Stinson}, {Sutter}, {Turchaninov}, {Tweed}, {Yepes}, \&
  {Zemp}}]{2011MNRAS.415.2293K}
{Knebe}, A., {et~al.} 2011, \mnras, 415, 2293

\bibitem[{{Kobayashi} \& {Nomoto}(2009)}]{2009ApJ...707.1466K}
{Kobayashi}, C., \& {Nomoto}, K. 2009, \apj, 707, 1466

\bibitem[{{Komatsu} {et~al.}(2011){Komatsu}, {Smith}, {Dunkley}, {Bennett},
  {Gold}, {Hinshaw}, {Jarosik}, {Larson}, {Nolta}, {Page}, {Spergel},
  {Halpern}, {Hill}, {Kogut}, {Limon}, {Meyer}, {Odegard}, {Tucker}, {Weiland},
  {Wollack}, \& {Wright}}]{2011ApJS..192...18K}
{Komatsu}, E., {et~al.} 2011, \apjs, 192, 18

\bibitem[{{Kroupa}(2002)}]{2002Sci...295...82K}
{Kroupa}, P. 2002, Science, 295, 82

\bibitem[{{Lada} {et~al.}(2010){Lada}, {Lombardi}, \&
  {Alves}}]{2010ApJ...724..687L}
{Lada}, C.~J., {Lombardi}, M., \& {Alves}, J.~F. 2010, \apj, 724, 687

\bibitem[{{Latif} {et~al.}(2012){Latif}, {Schleicher}, {Schmidt}, \&
  {Niemeyer}}]{2012arXiv1212.1619L}
{Latif}, M.~A., {Schleicher}, D.~R.~G., {Schmidt}, W., \& {Niemeyer}, J. 2012,
  ArXiv e-prints

\bibitem[{{Machacek} {et~al.}(2001){Machacek}, {Bryan}, \&
  {Abel}}]{2001ApJ...548..509M}
{Machacek}, M.~E., {Bryan}, G.~L., \& {Abel}, T. 2001, \apj, 548, 509

\bibitem[{{McBride} {et~al.}(2009){McBride}, {Fakhouri}, \&
  {Ma}}]{2009MNRAS.398.1858M}
{McBride}, J., {Fakhouri}, O., \& {Ma}, C.-P. 2009, \mnras, 398, 1858

\bibitem[{{McKee} \& {Tan}(2008)}]{2008ApJ...681..771M}
{McKee}, C.~F., \& {Tan}, J.~C. 2008, \apj, 681, 771

\bibitem[{{Mo} \& {White}(1996)}]{1996MNRAS.282..347M}
{Mo}, H.~J., \& {White}, S.~D.~M. 1996, \mnras, 282, 347

\bibitem[{{Nomoto} {et~al.}(2006){Nomoto}, {Tominaga}, {Umeda}, {Kobayashi}, \&
  {Maeda}}]{2006NuPhA.777..424N}
{Nomoto}, K., {Tominaga}, N., {Umeda}, H., {Kobayashi}, C., \& {Maeda}, K.
  2006, Nuclear Physics A, 777, 424

\bibitem[{{Norman} \& {Bryan}(1999)}]{1999ASSL..240...19N}
{Norman}, M.~L., \& {Bryan}, G.~L. 1999, in Astrophysics and Space Science
  Library, Vol. 240, Numerical Astrophysics, ed. S.~M. {Miyama}, K.~{Tomisaka},
  \& T.~{Hanawa}, 19

\bibitem[{{O'Shea} {et~al.}(2004){O'Shea}, {Bryan}, {Bordner}, {Norman},
  {Abel}, {Harkness}, \& {Kritsuk}}]{2004astro.ph..3044O}
{O'Shea}, B.~W., {Bryan}, G., {Bordner}, J., {Norman}, M.~L., {Abel}, T.,
  {Harkness}, R., \& {Kritsuk}, A. 2004, in Lecture Notes in Computational
  Science and Engineering, Vol.~41, Adaptive Mesh Refinement - Theory and
  Applications, ed. T.~{Plewa}, T.~{Linde}, \& V.~G. {Weirs}, 341

\bibitem[{{O'Shea} {et~al.}(2008){O'Shea}, {McKee}, {Heger}, \&
  {Abel}}]{2008AIPC..990D..13O}
{O'Shea}, B.~W., {McKee}, C.~F., {Heger}, A., \& {Abel}, T. 2008, in American
  Institute of Physics Conference Series, Vol. 990, First Stars III, ed. B.~W.
  {O'Shea} \& A.~{Heger}, D13

\bibitem[{{O'Shea} {et~al.}(2005){O'Shea}, {Nagamine}, {Springel}, {Hernquist},
  \& {Norman}}]{2005ApJS..160....1O}
{O'Shea}, B.~W., {Nagamine}, K., {Springel}, V., {Hernquist}, L., \& {Norman},
  M.~L. 2005, \apjs, 160, 1

\bibitem[{{O'Shea} \& {Norman}(2007)}]{2007ApJ...654...66O}
{O'Shea}, B.~W., \& {Norman}, M.~L. 2007, \apj, 654, 66

\bibitem[{{O'Shea} \& {Norman}(2008)}]{2008ApJ...673...14O}
---. 2008, \apj, 673, 14

\bibitem[{{Ricotti} {et~al.}(2002{\natexlab{a}}){Ricotti}, {Gnedin}, \&
  {Shull}}]{2002ApJ...575...33R}
{Ricotti}, M., {Gnedin}, N.~Y., \& {Shull}, J.~M. 2002{\natexlab{a}}, \apj,
  575, 33

\bibitem[{{Ricotti} {et~al.}(2002{\natexlab{b}}){Ricotti}, {Gnedin}, \&
  {Shull}}]{2002ApJ...575...49R}
---. 2002{\natexlab{b}}, \apj, 575, 49

\bibitem[{{Salpeter}(1955)}]{1955ApJ...121..161S}
{Salpeter}, E.~E. 1955, \apj, 121, 161

\bibitem[{{Schaerer}(2003)}]{2003A&A...397..527S}
{Schaerer}, D. 2003, \aap, 397, 527

\bibitem[{{Sheth} \& {Tormen}(1999)}]{1999MNRAS.308..119S}
{Sheth}, R.~K., \& {Tormen}, G. 1999, \mnras, 308, 119

\bibitem[{{Smith} {et~al.}(2009){Smith}, {Turk}, {Sigurdsson}, {O'Shea}, \&
  {Norman}}]{2009ApJ...691..441S}
{Smith}, B.~D., {Turk}, M.~J., {Sigurdsson}, S., {O'Shea}, B.~W., \& {Norman},
  M.~L. 2009, \apj, 691, 441

\bibitem[{{Smith} {et~al.}(2012){Smith}, {Iocco}, {Glover}, {Schleicher},
  {Klessen}, {Hirano}, \& {Yoshida}}]{2012ApJ...761..154S}
{Smith}, R.~J., {Iocco}, F., {Glover}, S.~C.~O., {Schleicher}, D.~R.~G.,
  {Klessen}, R.~S., {Hirano}, S., \& {Yoshida}, N. 2012, \apj, 761, 154

\bibitem[{{Stacy} \& {Bromm}(2007)}]{2007MNRAS.382..229S}
{Stacy}, A., \& {Bromm}, V. 2007, \mnras, 382, 229

\bibitem[{{Stacy} \& {Bromm}(2012)}]{2012arXiv1211.1889S}
---. 2012, ArXiv e-prints

\bibitem[{{Stacy} {et~al.}(2010){Stacy}, {Greif}, \&
  {Bromm}}]{2010MNRAS.403...45S}
{Stacy}, A., {Greif}, T.~H., \& {Bromm}, V. 2010, \mnras, 403, 45

\bibitem[{{Stacy} {et~al.}(2012){Stacy}, {Greif}, \&
  {Bromm}}]{2012MNRAS.422..290S}
---. 2012, \mnras, 422, 290

\bibitem[{{Tan} \& {McKee}(2004)}]{2004ApJ...603..383T}
{Tan}, J.~C., \& {McKee}, C.~F. 2004, \apj, 603, 383

\bibitem[{{Tegmark} {et~al.}(1997){Tegmark}, {Silk}, {Rees}, {Blanchard},
  {Abel}, \& {Palla}}]{1997ApJ...474....1T}
{Tegmark}, M., {Silk}, J., {Rees}, M.~J., {Blanchard}, A., {Abel}, T., \&
  {Palla}, F. 1997, \apj, 474, 1

\bibitem[{{Trenti} \& {Stiavelli}(2009)}]{2009ApJ...694..879T}
{Trenti}, M., \& {Stiavelli}, M. 2009, \apj, 694, 879

\bibitem[{{Tumlinson}(2006)}]{2006ApJ...641....1T}
{Tumlinson}, J. 2006, \apj, 641, 1

\bibitem[{{Tumlinson}(2010)}]{2010ApJ...708.1398T}
---. 2010, \apj, 708, 1398

\bibitem[{{Turk} {et~al.}(2009){Turk}, {Abel}, \&
  {O'Shea}}]{2009Sci...325..601T}
{Turk}, M.~J., {Abel}, T., \& {O'Shea}, B. 2009, Science, 325, 601

\bibitem[{{Turk} {et~al.}(2011{\natexlab{a}}){Turk}, {Clark}, {Glover},
  {Greif}, {Abel}, {Klessen}, \& {Bromm}}]{2011ApJ...726...55T}
{Turk}, M.~J., {Clark}, P., {Glover}, S.~C.~O., {Greif}, T.~H., {Abel}, T.,
  {Klessen}, R., \& {Bromm}, V. 2011{\natexlab{a}}, \apj, 726, 55

\bibitem[{{Turk} {et~al.}(2012){Turk}, {Oishi}, {Abel}, \&
  {Bryan}}]{2012ApJ...745..154T}
{Turk}, M.~J., {Oishi}, J.~S., {Abel}, T., \& {Bryan}, G.~L. 2012, \apj, 745,
  154

\bibitem[{{Turk} {et~al.}(2011{\natexlab{b}}){Turk}, {Smith}, {Oishi}, {Skory},
  {Skillman}, {Abel}, \& {Norman}}]{2011ApJS..192....9T}
{Turk}, M.~J., {Smith}, B.~D., {Oishi}, J.~S., {Skory}, S., {Skillman}, S.~W.,
  {Abel}, T., \& {Norman}, M.~L. 2011{\natexlab{b}}, \apjs, 192, 9

\bibitem[{{Warren} {et~al.}(2006){Warren}, {Abazajian}, {Holz}, \&
  {Teodoro}}]{2006ApJ...646..881W}
{Warren}, M.~S., {Abazajian}, K., {Holz}, D.~E., \& {Teodoro}, L. 2006, \apj,
  646, 881

\bibitem[{{Whalen} {et~al.}(2010){Whalen}, {Hueckstaedt}, \&
  {McConkie}}]{2010ApJ...712..101W}
{Whalen}, D., {Hueckstaedt}, R.~M., \& {McConkie}, T.~O. 2010, \apj, 712, 101

\bibitem[{{Whalen} {et~al.}(2008){Whalen}, {van Veelen}, {O'Shea}, \&
  {Norman}}]{2008ApJ...682...49W}
{Whalen}, D., {van Veelen}, B., {O'Shea}, B.~W., \& {Norman}, M.~L. 2008, \apj,
  682, 49

\bibitem[{{Wise} \& {Abel}(2005)}]{2005ApJ...629..615W}
{Wise}, J.~H., \& {Abel}, T. 2005, \apj, 629, 615

\bibitem[{{Wise} {et~al.}(2012{\natexlab{a}}){Wise}, {Abel}, {Turk}, {Norman},
  \& {Smith}}]{2012MNRAS.427..311W}
{Wise}, J.~H., {Abel}, T., {Turk}, M.~J., {Norman}, M.~L., \& {Smith}, B.~D.
  2012{\natexlab{a}}, \mnras, 427, 311

\bibitem[{{Wise} {et~al.}(2012{\natexlab{b}}){Wise}, {Turk}, {Norman}, \&
  {Abel}}]{2012ApJ...745...50W}
{Wise}, J.~H., {Turk}, M.~J., {Norman}, M.~L., \& {Abel}, T.
  2012{\natexlab{b}}, \apj, 745, 50

\bibitem[{{Wolcott-Green} \& {Haiman}(2011)}]{2011MNRAS.412.2603W}
{Wolcott-Green}, J., \& {Haiman}, Z. 2011, \mnras, 412, 2603

\bibitem[{{Wolcott-Green} {et~al.}(2011){Wolcott-Green}, {Haiman}, \&
  {Bryan}}]{2011MNRAS.418..838W}
{Wolcott-Green}, J., {Haiman}, Z., \& {Bryan}, G.~L. 2011, \mnras, 418, 838

\bibitem[{{Yoshida} {et~al.}(2003){Yoshida}, {Abel}, {Hernquist}, \&
  {Sugiyama}}]{2003ApJ...592..645Y}
{Yoshida}, N., {Abel}, T., {Hernquist}, L., \& {Sugiyama}, N. 2003, \apj, 592,
  645

\end{thebibliography}

\end{document}